\input{aipcheck}

\documentclass[
    ,final            % use final for the camera ready runs
  ,numberedheadings % uncomment this option for numbered sections
  ]
  {aipproc}

\usepackage{bm}

\layoutstyle{6x9}

\newcommand{\msun}{M_{\odot}}
\newcommand{\rsun}{R_{\odot}}
\newcommand{\lsun}{L_{\odot}}
\newcommand{\eff}{\epsilon_{\rm ff}}
\newcommand{\avir}{\alpha_{\rm vir}}
\newcommand{\vecr}{\mathbf{r}}
\newcommand{\vecv}{\mathbf{v}}
\newcommand{\vecx}{\mathbf{x}}

\newcommand{\vecB}{\mathbf{B}}

\newcommand{\vecS}{\mathbf{S}}
\newcommand{\vecI}{\mathbf{I}}
\newcommand{\vecPi}{\Pi}
\newcommand{\vecT}{\mathbf{T}}

\newcommand{\ltsim}{\protect\raisebox{-0.5ex}{$\:\stackrel{\textstyle <}
        {\sim}\:$}}
\newcommand{\gtsim}{\protect\raisebox{-0.5ex}{$\:\stackrel{\textstyle >}
        {\sim}\:$}}

\begin{document}

\title{Star Formation in Molecular Clouds}

\classification{<Replace this text with PACS numbers; choose from this list:
                \texttt{http://www.aip..org/pacs/index.html}>}

\author{Mark R.~Krumholz}{
  address={Department of Astronomy \& Astrophysics, University of California, Santa Cruz}
}

\begin{abstract}
Star formation is one of the least understood processes in cosmic evolution. It is difficult to formulate a general theory for star formation in part because of the wide range of physical processes involved. The interstellar gas out of which stars form is a supersonically turbulent plasma governed by magnetohydrodynamics. This is hard enough by itself, since we do not understand even subsonic hydrodynamic turbulence very well, let alone supersonic non-ideal MHD turbulence. However, the behavior of star-forming clouds in the ISM is also obviously influenced by gravity, which adds complexity, and by both continuum and line radiative processes. Finally, the behavior of star-forming clouds is influenced by a wide variety of chemical processes, including formation and destruction of molecules and dust grains (which changes the thermodynamic behavior of the gas) and changes in ionization state (which alter how strongly the gas couples to magnetic fields). As a result of these complexities, there is nothing like a generally agreed-upon theory of star formation, as there is for stellar structure. Instead, we are forced to take a much more phenomenological approach. These notes provide an introduction to our current thinking about how star formation works.
\end{abstract}

\maketitle

%%%%%%%%%%%%%%%%%%%%%%%%%%%%%%%%%%%%%%%%%%%%
%% MAINMATTER
%%%%%%%%%%%%%%%%%%%%%%%%%%%%%%%%%%%%%%%%%%%%

\section{Forward}

These proceedings are based on a series of lectures given at the XV$^{\rm th}$ Special Courses of the National Observatory of Rio de Janeiro, the overall goal of which is to provide a crash course in star formation for beginning graduate students or advanced undergraduates. Because this text is meant to be pedagogic, it is generally much more explicit about the algebra and methods behind calculations than a standard journal article. Due to limits of space, these notes are necessarily incomplete, and they are biased in places by the author's opinions (both about what is interesting and about what is correct). Caveat lector. For a more comprehensive overview of the field, the best source is the recent review by McKee \& Ostriker \cite{mckee07b}. A much more extensive pedagogic introduction, which is unfortunately also fairly dated at this point, may be found in the textbook by Stahler \& Palla \cite{stahler05a}. Some of the material included in these lectures also covers basics of radiative transfer and fluid dynamics, and students looking for more information on these topics may consult standard textbooks such as Rybicki \& Lightman (for radiation) and Shu (for both fluid dynamics and radiation).

In these proceedings, each section corresponds to a single lecture. The first section discusses how we observe star-forming clouds, also known as molecular clouds (at least in the Milky Way) and determine their properties. The second begins to investigate the physical processes that govern the behavior of the clouds. In the third we discuss why molecular clouds collapse and what happens when they do, including the critical problems of how energy, angular momentum, and magnetic flux are transported. Finally, the fourth section focuses on perhaps the two major unsolved problems of star formation: the star formation rate and the initial mass function. It is in this section that the reader should be particularly aware of author biases, since the material in the first three sections is generally (though not always) non-controversial, while the material in the final section is far from it.

\section{Observing Star-Forming Clouds}

\subsection{Observational Techniques}

We will begin with a discussion of the important observational techniques that we use to obtain information about the star-forming ISM. This will naturally lead us to review some of the important radiative transfer physics that we need to keep in mind to understand the observations. Because the interstellar clouds that form stars are generally cold, most (but not all) of these techniques require on infrared, sub-millimeter, and radio observations.

\subsubsection{The Problem of H$_2$}
  
Hydrogen is the most abundant element, and when it is in the form of free atomic hydrogen, it is relatively easy to observe. Hydrogen atoms have a hyperfine transition at 21 cm (1.4 GHz), associated with a transition from a state in which the spin of the electron is parallel to that of the proton to a state where it is anti-parallel. The energy associated with this transition is $\ll 1$ K, so even in cold regions it can be excited. This line is seen in the Milky Way and in many nearby galaxies.
  
However, at the high densities where stars form, hydrogen tends to be molecular rather than atomic, and H$_2$ is extremely hard to observe directly. To understand why, we must review the quantum structure of H$_2$. A diatomic molecule like H$_2$ has three types of excitation: electronic (corresponding to excitations of one or more of the electrons), vibrational (corresponding to vibrational motion of the two nuclei), and rotational (corresponding to rotation of the two nuclei about the center of mass). Generally electronic excitations are highest in energy scale, vibrational are next, and rotational are the lowest in energy.

For H$_2$, the first excited state, the $J=1$ rotational state, is $100-200$ K above the ground state.
This energy gap between the ground state and the first excited state is far larger than for any other simple molecule, and the underlying reason for this large energy is the low mass of hydrogen. For a quantum oscillator or rotor the level spacing varies with reduced mass as $m^{-1/2}$. Since the dense ISM where molecules form is often also cold, $T\sim 10$ K (as we discuss later), almost no molecules will be in this excited state. However, it gets even worse: H$_2$ is a homonuclear molecule, which means that it has zero electric dipole moment. As a result, electric dipole transitions do not occur, and radiative transitions that change $J$ by 1 are electric dipoles. This means that there is no $J=1\rightarrow 0$ emission. Instead, the lowest-lying transition is the $J=2\rightarrow 0$ quadrupole. This is very weak, because it's a quadrupole. More importantly, however, the $J=2$ state is 511 K above the ground state. This means that, for a population in equilibrium at a temperature of 10 K, the fraction of molecules in the $J=2$ state is $\sim e^{-10/500} \approx 10^{-22}$! In effect, in a molecular cloud there are simply no H$_2$ molecules in states capable of emitting.
  
The conclusion of this analysis is that, for typical conditions in star-forming clouds, we cannot observe the most abundant species, H$_2$, in emission. Instead, we are forced to observe proxies instead. (One can observe H$_2$ in absorption against background sources, but this is possible only in special circumstances.)

\subsubsection{Observing the Dust}
  
One proxy we can use, which is perhaps the most straightforward conceptually, is dust.  Interstellar gas clouds are always mixed with dust, and the dust grains emit thermal radiation which we can observe. They also absorb background starlight, and we observe that absorption too. The advantage of dust grains is that, since they are solid particles, the can absorb or emit continuum radiation, which the gas cannot. Consider a cloud of gas of mass density $\rho$ mixed with dust grains at a temperature $T$.  The gas-dust mixture has a specific opacity $\kappa_{\nu}$ to radiation at frequency $\nu$. Although the vast majority of the mass is in gas rather than dust, the opacity will be almost entirely due to the dust grains except for frequencies that happen to match the resonant absorption frequencies of atoms and molecules in the gas. 
 
Radiation passing through the cloud is governed by the equation of radiative transfer:
\begin{equation}
\frac{dI_\nu}{ds} = j_{\nu} - \kappa_{\nu} I_\nu,
\end{equation}
where $I_\nu$ is the radiation intensity, and we integrate along a path through the cloud. The emissivity for gas of opacity $\kappa_{\nu}$ that is in local thermodynamic equilibrium (LTE) is $j_{\nu} = \kappa_{\nu} \rho B_{\nu}(T)$, where $j_{\nu}$ has units of erg s$^{-1}$ cm$^{-3}$ sr$^{-1}$ Hz$^{-1}$, i.e.\ it describes the number of ergs emitted in 1 second by 1 cm$^3$ of gas into a solid angle of 1 sr in a frequency range of 1 Hz, and
\begin{equation}
B_{\nu}(T) = \frac{2 h\nu^3}{c^2} \frac{1}{e^{h\nu/kT}-1}
\end{equation}
is the Planck function.

Generally we look for emission at submillimeter wavelengths, and for absorption at near infrared wavelengths. In the sub-mm typical opacities are $\kappa_{\nu} \sim 0.01$ cm$^{2}$ g$^{-1}$. Since essentially no interstellar cloud has a surface density $> 100$ g cm$^{-2}$, absorption of radiation from the back of the cloud by gas in front of it is completely negligible. Thus, we can set $\kappa_\nu I_\nu$ to zero in the transfer equation, and integrate trivially:
\begin{equation}
I_{\nu} = \int j_{\nu} ds = \Sigma \kappa_{\nu} B_{\nu}(T) = \tau_{\nu} B_{\nu}(T)
\end{equation}
where $\Sigma=\int \rho ds$ is the surface density of the cloud and $\tau_{\nu} = \Sigma \kappa_{\nu}$ is the optical depth of the cloud at frequency $\nu$.
Thus if we observe the intensity of emission from dust grains in a cloud, we determine the product of the optical depth and the Planck function, which is determined solely by the observing frequency and the gas temperature. If we know the temperature and the properties of the dust grains, we can therefore determine the column density of the gas in the cloud in each telescope beam.

Conversely, if we are looking for absorption in the near-IR, we have a background star that emits light that enters the cloud with intensity $I_{\nu,0}$. The cloud itself emits negligibly in the near-IR, because $h\nu \gg kT$, so the exponential factor in the denominator of the Planck function is huge. Thus we can drop the $j_\nu$ term in the transfer equation, and the solution is again trivial:
\begin{equation}
I_{\nu} = I_{\nu,0} e^{-\tau_\nu}.
\end{equation}
By measuring the optical depth at several frequencies, and knowing the intrinsic frequency-dependence of $I_{\nu,0}$ for stars, we can figure out the optical depth and thus the column density.

\begin{figure}
\includegraphics[height=.3\textheight]{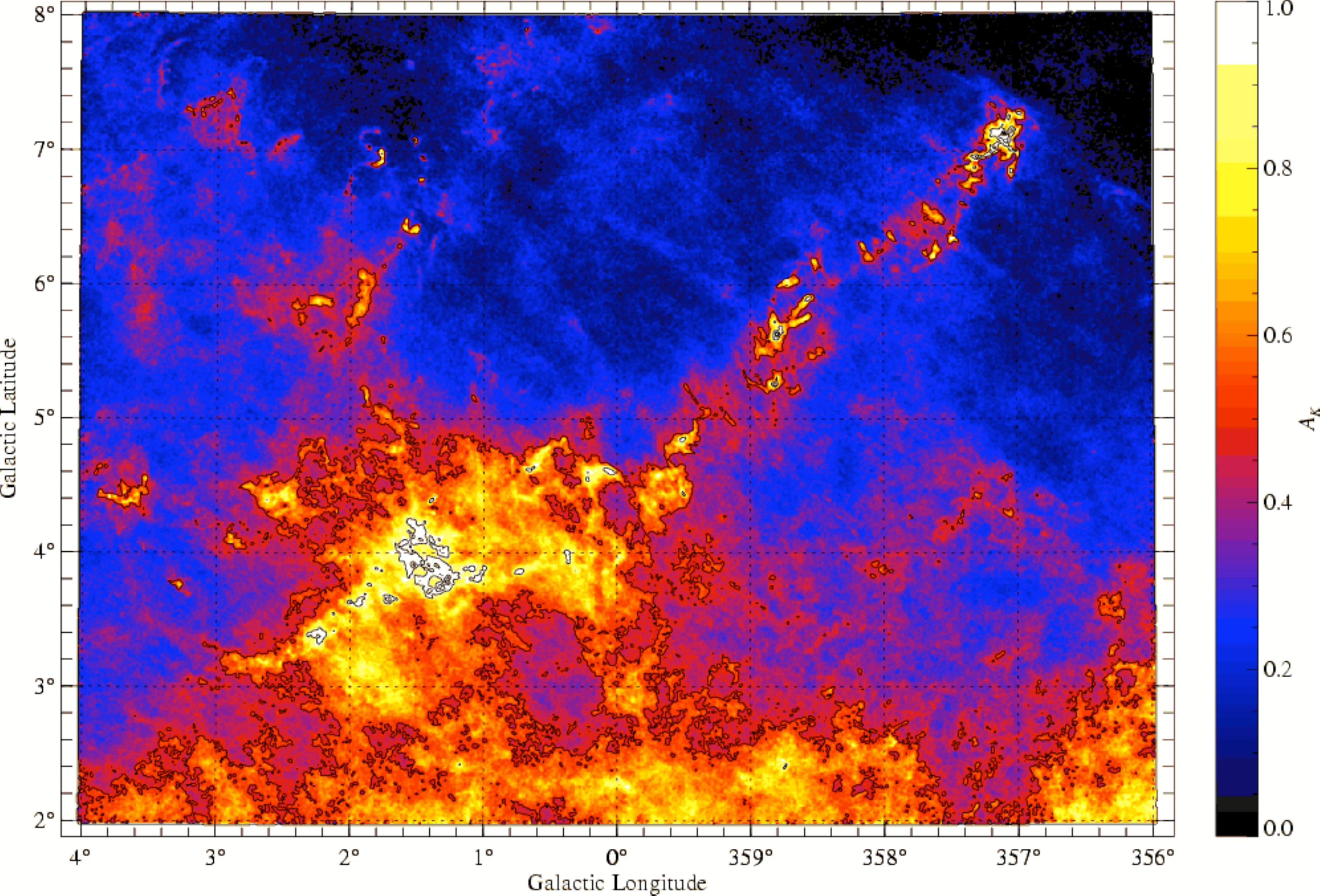}
  \caption{A map of the Pipe Nebula obtained with near-infrared absorption measurements. Color indicates visual extinction, which is proportional to column density. Reprinted with permission from \citep{lombardi06a}.
  \label{fig:pipenebula}}
\end{figure}

These mapping techniques allow us to obtain extremely detailed maps of nearby molecular clouds. Figure \ref{fig:pipenebula} shows a spectacular example. Unfortunately, these techniques are generally not usable for extragalactic observations. The resolution and sensitivity of sub-mm telescopes is not sufficient to allow us to see individual clouds in emission, and the problem of knowing which stars are behind or in front of a given gas cloud at extragalactic distances prevents us from making good measurements in absorption. Both of these limitations may be eased by future telescopes, but for now dust observations of individual clouds are generally limited to the Milky Way.

\subsubsection{Molecular lines}

Much of what we know about star forming gas comes from observations of molecular line emission. These are usually the most complex measurements in terms of the modeling and required to understand them. However, they are also by far the richest in terms of the information they provide. They are also among the most sensitive, since the lines can be very bright compared to continuum emission. Indeed, almost everything we know about giant molecular clouds outside of our own galaxy comes from studying emission in the rotational lines of the CO molecule. The CO molecule, since it is much more massive than H$_2$, has its lowest rotational state only $5.5$ K above ground, low enough to be excited even at GMC temperatures. Since C and O are two of the most common elements in the ISM beyond H and He, CO molecules are abundant and the lines are bright.

\subsubsection{Two-Level Atoms}

The simplest line-emitting system is an atom or molecule with exactly two energy states, but this example contains most of the concepts we will need. We'll explore how that works first, then consider more complex, realistic molecules. Consider an atom or molecule of species $X$ with two states that are separated by an energy $E$. Suppose we have a gas of such particles with number density $n_X$ at temperature $T$. The number density of atoms in the ground state is $n_0$ and the number density in the excited state is $n_1$. At first suppose that this system does not radiate. In this case collisions between the atoms will eventually bring the two energy levels into thermal equilibrium. In that case, what are $n_0$ and $n_1$?

They just follow a Boltzmann distribution, so $n_1/n_0 = e^{-E/kT}$, and thus we have $n_0 = n_X /Z$ and $n_1 = n_X e^{-E/kT}/Z$, where $Z=1+e^{-E/kT}$ is the partition function. Gas with such a distribution of level populations is said to be in LTE. Now let's consider radiative transitions between these states. There are three processes: spontaneous emission, stimulated emission, and absorption, which are described by the three Einstein coefficients. For simplicity we'll start by neglecting all but spontaneous emission. This is sometimes a good approximation in the interstellar medium, since in many cases the ambient radiation field is too weak for stimulated emission or absorption to be important.

A particle in the excited state can spontaneously emit a photon and decay to the ground state. The rate at which this happens is described by the Einstein coefficient $A_{10}$, which has units of s$^{-1}$. Its meaning is simply that a population of $n_1$ atoms in the excited state will decay to the ground state by spontaneous emission at a rate 
\begin{equation}
\left(\frac{dn_1}{dt}\right)_{\rm spon.~emis.} = -A_{10} n_1
\end{equation}
atoms per cm$^3$ per s, or equivalently that the $e$-folding time for decay is $1/A_{10}$ seconds.

For the molecules we'll be spending most of our time talking about, decay times are typically at most a few centuries, which is short compared to pretty much any time scale associated with star formation. Thus if spontaneous emission were the only process at work, all molecules would quickly decay to the ground state and we wouldn't see any emission. However, in the dense interstellar environments where stars form, collisions occur frequently enough to create a population of excited molecules. Of course collisions involving excited molecules can also cause de-excitation, with the excess energy going into recoil rather than into a photon.

Since hydrogen molecules are almost always the most abundant species in the dense regions we're going to think about, with helium second, we can generally only consider collisions between our two-level atom and those partners. For the purposes of this exercise, we'll ignore everything but H$_2$. The rate at which collisions cause transitions between states is a horrible quantum mechanical problem. We cannot even confidently calculate the energy levels of single isolated molecules except in the simplest cases, let alone the interactions between two colliding ones at arbitrary velocities and relative orientations. Exact calculations of collision rates are generally impossible. Instead, we either make due with approximations (at worst), or we try to make laboratory measurements. Things are bad enough that, for example, we often assume that the rates for collisions with H$_2$ molecules and He atoms are related by a constant factor. 

Fortunately, as astronomers we generally leave these problems to chemists, and instead do what we always do: hide our ignorance behind a parameter. We let the rate at which collisions between species X and H$_2$ molecules induce transitions from the ground state to the excited state be
\begin{equation}
\left(\frac{dn_1}{dt}\right)_{\rm coll.~exc.} = \gamma_{01} n_0 n,
\end{equation}
where $n$ is the number density of H$_2$ molecules ({\it not} the number density of species X) and $\gamma_{01}$ has units of cm$^3$ s$^{-1}$. In general $\gamma_{01}$ will be a function of the gas kinetic temperature $T$, but not of $n$ (unless $n$ is so high that three-body processes start to become important, which is almost never the case in the ISM). The corresponding rate coefficient for collisional de-excitation is $\gamma_{10}$, and the collisional de-excitation rate is
\begin{equation}
\left(\frac{dn_1}{dt}\right)_{\rm coll.~de-exc.} = -\gamma_{10} n_1 n.
\end{equation}
Collections of collision rate coefficients for common molecules can be found in the extremely useful Leiden Atomic and Molecular Database\footnote{\url{http://www.strw.leidenuniv.nl/~moldata/}} \citep{schoier05a}.

A little thought will convince you that $\gamma_{01}$ and $\gamma_{10}$ must have a specific relationship. Consider an extremely optically thick region where so few photons escape that radiative processes are not significant. If the gas is in equilibrium then we have
\begin{eqnarray}
\frac{dn_1}{dt} = \left(\frac{dn_1}{dt}\right)_{\rm coll.~exc.} + \left(\frac{dn_1}{dt}\right)_{\rm coll.~de-exc.} & = & 0 \\
n (\gamma_{01} n_0 - \gamma_{10} n_1) & = & 0.
\end{eqnarray}
However, we also know that the equilibrium distribution is a Boltzmann distribution, so $n_1/n_0 = e^{-E/kT}$. Thus we have
\begin{eqnarray}
n n_0 (\gamma_{01} - \gamma_{10} e^{-E/kT}) & = & 0 \\
\gamma_{01} & = & \gamma_{10} e^{-E/kT}.
\end{eqnarray}
This argument applies equally well between a pair of levels even for a complicated molecule with many levels instead of just 2. Thus, we only need to know the rate of collisional excitation or de-excitation between any two levels to know the reverse rate.

We are now in a position to write down the full equations of statistical equilibrium for the two-level system. In so doing, we will see that we can immediately use line emission to learn a great deal about the density of gas. In equilibrium we have
\begin{eqnarray}
\frac{dn_1}{dt} & = & 0 \\
n_1 A_{10} + n n_1 \gamma_{10} -n n_0 \gamma_{01} & = & 0 \\
\frac{n_1}{n_0} \left(A_{10} + \gamma_{10}n\right) - \gamma_{01} n & = & 0\\
\frac{n_1}{n_0} & = & \frac{\gamma_{01} n}{A_{10}+\gamma_{10} n}\\
& = & e^{-E/kT} \frac{1}{1+A_{10}/(\gamma_{10} n)}
\end{eqnarray}
This physical meaning of this expression is clear. If radiation is negligible compared to collisions, i.e.\ $A_{10} \ll \gamma_{10} n$, then the ratio of level populations approaches the Boltzmann ratio $e^{-E/kT}$. As radiation becomes more important, i.e.\ $A_{10}/(\gamma_{10} n)$ get larger, the fraction in the upper level drops -- the level population is sub-thermal. This is because radiative decays remove molecules from the upper state much faster than collisions re-populate it.

Since the collision rate depends on density and the radiative decay rate does not, the balance between these two processes depends on density. This make it convenient to introduce a critical density $n_{\rm crit}$, defined by n$_{\rm crit} = A_{10} / \gamma_{10}$, so that
\begin{equation}
\frac{n_1}{n_0} = e^{-E/kT} \frac{1}{1+n_{\rm crit}/n}.
\end{equation}
At densities much larger than $n_{\rm crit}$, we expect the level population to be close to the Boltzmann value, and at densities much smaller than $n_{\rm crit}$ we expect the upper state to be under-populated relative to Boltzmann. $n_{\rm crit}$ itself is simply the density at which radiative and collisional de-excitations out of the upper state occur at the same rate.

For real molecules or atoms with more than two states, the critical density for state $i$ can be generalized to
\begin{equation}
n_{{\rm crit},i} = \frac{\sum_{j<i} A_{ij}}{\sum_{j<i} k_{ij}},
\end{equation}
i.e.\ the critical density is simply the sum of the Einstein $A$'s for all levels less than $i$, divided by the sum of the collision rate coefficients for transitions from level $i$ to all levels less than $i$. The condition for equilibrium is
\begin{equation}
n_i = \left(\frac{1}{1+n_{{\rm crit},i}/n}\right) \frac{\sum_{j<i} n_j k_{ji}}{\sum_{j<i} k_{ij}}
\end{equation}
This is a series of linear equations (one for each level $i$) that can be solved to give the level populations. We could write down an exact solution in terms of a matrix inversion, but it's more illuminating just to notice how the solution will have to behave. For $n \gg n_{{\rm crit},i}$, the leading term in parentheses goes to unity, and the relationships between the different level populations $n_i$ are just determined by the collision rate coefficients $k_{ij}$ -- the Einstein coefficient drops out of the problem. In this case, the level populations go to the Boltzmann distribution. For $n \ll n_{{\rm crit},i}$, the leading term in parentheses is smaller than unity, and higher levels are underpopulated relative to the Boltzmann distribution. Thus the behavior is qualitatively similar to the two-level atom.

\subsection{Molecular Cloud Properties from Molecular Lines}

Molecular lines, as we have seen, are a rather complicated way to observe things, since the emission we get out depends on many factors. However, we can turn this to our advantage. The complexity of the molecular line emission process can be exploited to tell us all sorts of things about molecular clouds. Indeed, they form the basis of most of our knowledge of cloud properties. For the rest of this section we'll mostly go back to our two-level particle for simplicity, since the procedures for multi-level particles are analogous but more mathematically cumbersome.

\subsubsection{Density Inference}

First of all, let's consider the rate of energy emission per molecule from a molecular line. This is easy once we know the level population:
\begin{eqnarray}
\frac{\mathcal{L}}{n_X} & = & \frac{E A_{10} n_1}{n_X} \\
& = & E A_{10} \frac{n_1}{n_0+n_1} \\
& = & E A_{10} \frac{n_1/n_0}{1+n_1/n_0} \\
& = & E A_{10} \frac{e^{-E/kT}}{1+e^{-E/kT}+n_{\rm crit}/n} \\
& = & E A_{10} \frac{e^{-E/kT}}{Z+n_{\rm crit}/n},
\end{eqnarray}
where again $Z$ is the partition function. It is instructive to think about how this behaves in the limiting cases $n \ll n_{\rm crit}$ and $n\gg n_{\rm crit}$. 
In the limit $n\gg n_{\rm crit}$, the partition function $Z$ dominates the denominator, and we get $\mathcal{L}/n_X = E A_{10} e^{-E/kT}{Z}$. This is just the energy per spontaneous emission times the spontaneous emission rate times the fraction of the population in the upper state when the gas is in statistical equilibrium. This is density-independent, so this means that at high density you just get a fixed amount of emission per molecule of the emitting species. The total luminosity is just proportional to the number of emitting molecules.

\begin{figure}
\includegraphics[height=.3\textheight]{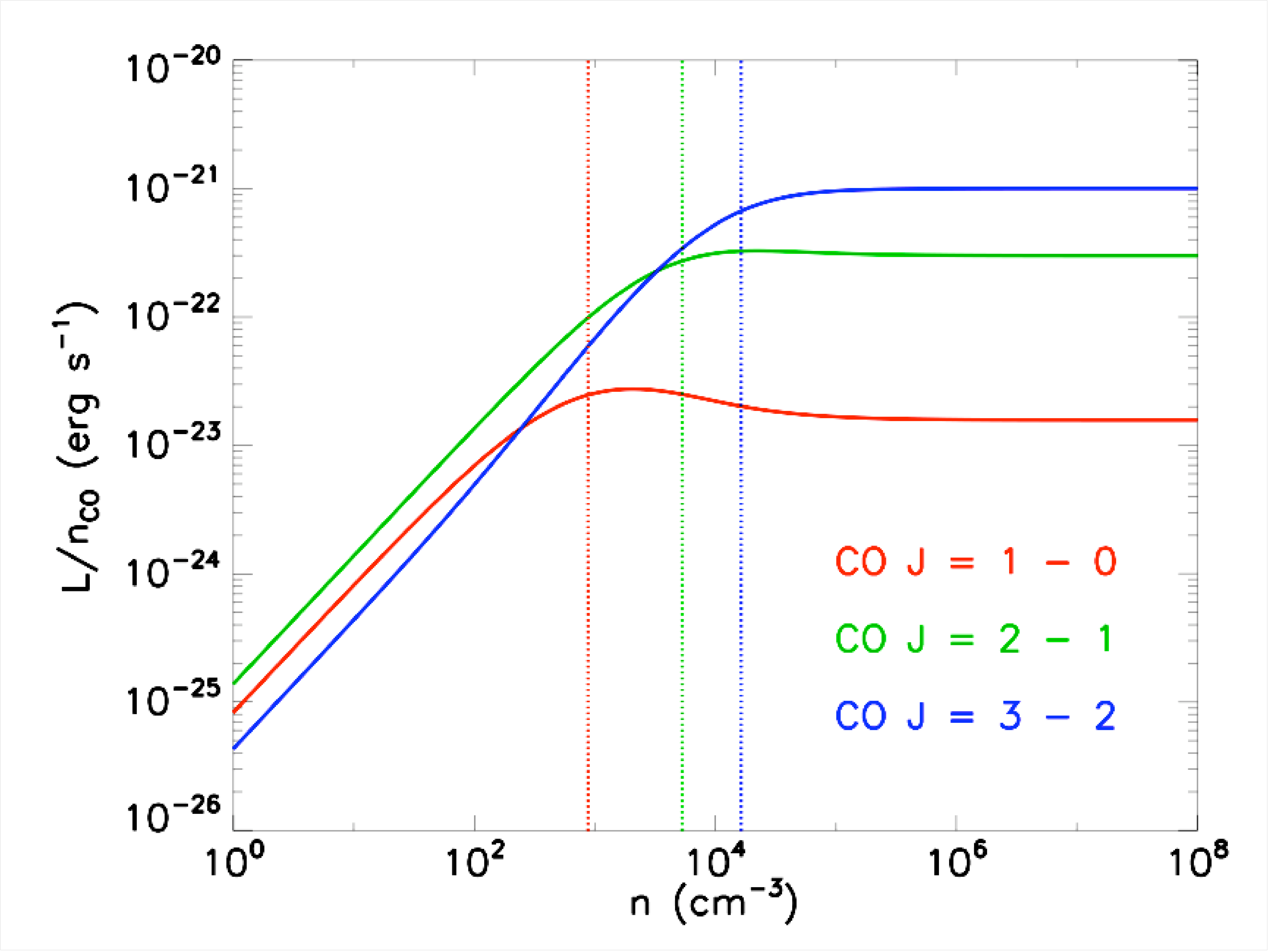}
  \caption{Emitted luminosity per CO molecule as a function of H$_2$ density $n$, for the $J=1\rightarrow 0$, $J=2\rightarrow 1$, and $J=3\rightarrow 2$ transitions (red, green, and blue, respectively), computed neglecting radiative trapping and assuming statistical equilibrium. The calculation assumes a temperature of 10 K, and uses the collision rate coefficients given in the Leiden Atomic and Molecular Database \cite{schoier05a}. The dashed vertical lines indicate the critical density for each transition. This simple calculation is done using the method described in the text, which is for pure molecular hydrogen. Including He would change these numbers very slightly.
  \label{fig:colum}}
\end{figure}

For $n \ll n_{\rm crit}$, the second term dominates the denominator, and we get
\begin{equation}
\frac{\mathcal{L}}{n_X} \approx E A_{10} e^{-E/kT} \frac{n}{n_{\rm crit}}.
\end{equation}
Thus at low density each molecule contributes an amount of light that is proportional to the ratio of density to critical density. Note that this is the ratio of collision partners, i.e.\ of H$_2$, rather than the density of emitting molecules. The total luminosity varies as this ratio times the number of emitting molecules. Figure \ref{fig:colum} shows an example of this behavior for the first three rotational transitions of CO.

The practical effect of this is that, at densities below the critical density, emission from a molecular line is often unobservably small. Even if it can be observed, emission from gas below the critical density is likely to be dwarfed by emission from gas at or above the critical density. This means that observing molecular lines often immediately tells us about the volume density of the gas! In effect, a measurement of the luminosity of a particular line tells us something like the total mass of gas that is dense enough to excite that line. If we observe a cloud in many different molecular lines with different critical densities, we can deduce the density distribution within that cloud. Analysis of this sort indicates that the bulk of the material in molecular clouds is at densities $n\sim 100$ cm$^{-3}$, while small amounts of mass reach much higher densities.

As a caution I should mention that this is computed for optically thin emission. If the line is optically thick, you can no longer ignore stimulated emission and absorption processes, and not all emitted photons will escape from the cloud. CO is usually optically thick. The effect of optical thickness is to reduce the effective critical density. This is because trapping of photons within the cloud means that not every spontaneously-emitted photon escapes the cloud, which has an effect like like lowering the Einstein $A$.

\subsubsection{Velocity and temperature inference}

We can also use molecular lines to infer the velocity and temperature structure of gas if the line in question is optically thin, meaning that we can neglect absorption. For an optically thin line, the width of the line is determined primarily by the velocity distribution of the emitting molecules. The physics here is extremely simple. Suppose we have gas along our line of sight with a velocity distribution $\psi(v)$, i.e.\ the fraction of gas with velocities between $v$ and $v+dv$ is $\psi(v) dv$, and $\int_{-\infty}^{\infty} \psi(v) \, dv = 0$. For an optically thin line, in the limit where natural and pressure-broadening of lines is negligible, we can think of emission producing a delta function in frequency in the rest frame of the gas. There is a one-to-one mapping between velocity and frequency. Thus emission from gas moving at a frequency $v$ relative to us along our line of sight produces emission at a frequency 
\begin{equation}
\nu \approx \nu_0 \left(1 - \frac{v}{c}\right),
\end{equation}
where $\nu_0$ is the central frequency of the line in the molecule's rest frame, and we assume $v/c \ll 1$.

In this case the line profile is described trivially by 
\begin{equation}
\phi(\nu)=\psi\left(c\left[1-\frac{\nu}{\nu_0}\right]\right).
\end{equation}
We can measure $\phi(\nu)$ directly, and this immediately tells us the velocity distribution $\psi(v)$. In general the velocity distribution of the gas $\psi(v)$ is produced by a combination of thermal and non-thermal motions. Thermal motions arise from the Maxwellian velocity distribution of the gas, and produce a Maxwellian profile $\phi(\nu)\propto e^{-(\nu-\nu_{\rm cen})^2/\sigma_\nu^2}$. Here $\nu_{\rm cen}$ is the central frequency of the line, which is $\nu_{\rm cen} = \nu_0 (1 - \bar{v}/c)$, where $\bar{v}$ is the mean velocity of the gas along our line of sight. The width is $\sigma_\nu = \sqrt{kT/\mu}/c$, where $T$ is the gas temperature and $\mu$ is the mean mass of the emitting molecule. This is just the 1D Maxwellian distribution.

Non-thermal motions involve bulk flows of the gas, and can produce a variety of velocity distributions depending how the cloud is moving. Unfortunately even complicated motions often produce distributions that look something like Maxwellian distributions, just because of the central limit theorem: if you throw together a lot of random junk, the result is usually a Gaussian distribution.

Determining whether a given line profile reflects predominantly thermal or non-thermal motion requires that we have a way of estimating the temperature independently. This can often be done by observing multiple lines of the same species. Our expression
\begin{equation}
\frac{\mathcal{L}}{n_X} = E A_{10} \frac{e^{-E/kT}}{Z + n_{\rm crit}/n}
\end{equation}
shows that the luminosity of a particular optically thin line is a function of the temperature $T$, the density $n$, and the number density of emitting molecules $n_X$. If we observe three transitions of the same molecule, then we have three equations in three unknowns and we can solve for $n$, $n_X$, and $T$ independently. Certain molecules, because of their level structures, make this technique particularly clean. The most famous example of this is ammonia, NH$_3$.

Measurements of this sort show that typical molecular clouds have velocity dispersions of several km s$^{-1}$, but very low temperatures of only $\sim 10$ K. This is significant because the sound speed for H$_2$ molecules at 10 K is $c_s = \sqrt{k T/m_{\rm H_2}} \approx 0.2$ km s$^{-1}$. Thus the observed linewidths indicate that the typical velocities of material inside a GMC are supersonic by factors of $\sim 10$. This has important implications that we will explore below.

\subsubsection{Mass inference}

The last thing we routinely infer from line observations is total masses of clouds. In this case we usually want to pick a line that is quite optically thick, such as CO $J=1\rightarrow 0$. Other commonly-used lines include CO $J=2\rightarrow 1$, $^{13}$CO $J=1\rightarrow 0$ and $J=2\rightarrow 1$, and HCN $J=1\rightarrow 0$. The main motivation for using an optically thick line is that these tend to be nice and bright, so they're observable in external galaxies, or at long distances within our galaxy.

The challenge for an optically thick line is how to infer a mass, given that we're really only seeing the surface of a cloud. At first blush this shouldn't be possible -- after all, I cannot infer how thick a wall is by seeing its surface. The reason it is possible is that molecular clouds are not like walls. Even at their surfaces they carry information about their full mass. To see why this is, consider optically thick line emission from a cloud of mass $M$ and radius $R$ at temperature $T$. The mean column density is $N= M/(\mu \pi R^2)$, where $\mu=3.9\times 10^{-24}$ g is the mass per H$_2$ molecule. 

Suppose this cloud is in virial balance between kinetic energy and gravity, so that its kinetic energy is half its potential energy (we'll discuss this more in the context of the virial theorem in the next section). The gravitational-self energy is $\mathcal{W}=-a GM^2/R$, where $a$ is a constant of order unity that depends on the cloud's geometry and internal mass distribution. For a uniform sphere $a=3/5$. The kinetic energy is $\mathcal{T}=(3/2)M\sigma_{\rm 1D}^2$, where $\sigma_{\rm 1D}$ is the one dimensional velocity dispersion, including both thermal and non-thermal components. We define the virial ratio as
\begin{equation}
\alpha_{\rm vir} =  \frac{5\sigma_{\rm 1D}^2 R}{GM},
\end{equation}
For a uniform sphere, which has $a=3/5$, this definition implies $\alpha_{\rm vir}=2\mathcal{T}/|\mathcal{W}|$. Thus $\alpha_{\rm vir}=1$ corresponds to the ratio of kinetic to gravitational energy in a uniform sphere of gas in virial equilibrium between internal motions and gravity. In general we expect that $\avir\approx 1$ in any object supported primarily by internal turbulent motion, even if its mass distribution is not uniform. Re-arranging this definition, we have
\begin{equation}
\sigma_{\rm 1D} = \sqrt{\left(\frac{\avir}{5}\right)\frac{GM}{R}}.
\end{equation}

To see why this is relevant for the line emission, consider the total frequency-integrated intensity that the line will emit. The emission will be dominated by gas with a density above the critical density, for the reason we just discussed. This gas is close to LTE, so its emissivity is given by the Planck function times its opacity. In this case the solution to the transfer equation is
\begin{equation}
I_{\nu} = \left(1-e^{-\tau_{\nu}}\right) B_{\nu}(T),
\end{equation}
so integrating over frequency we get
\begin{equation}
\int I_{\nu}\,d\nu = \int \left(1-e^{-\tau_{\nu}}\right) B_{\nu}(T)\,d\nu.
\end{equation}
By assumption the optical depth at line center is $\tau_{\nu_0}\gg 1$, and for a Gaussian line profile the optical depth at frequency $\nu$ is
\begin{equation}
\tau_{\nu} =\tau_{\nu_0} \exp\left[-\frac{(\nu-\nu_0)^2}{2 (\nu_0 \sigma_{\rm 1D}/c)^2}\right]
\end{equation}
Since the integrated intensity depends on the integral of $\tau_{\nu}$ over frequency, and the frequency-dependence of $\tau_{\nu}$ is determined by $\sigma_{\rm 1D}$, we therefore expect that the integrated intensity will depend on $\sigma_{\rm 1D}$. 

To get a sense of how this dependence will work, let us adopt a very simplified yet schematically correct form for $\tau_{\nu}$. We will take the opacity to be a step function, which is infinite near line center and drops sharply to 0 far from line center. The frequency at which this transition happens will be set by the condition $\tau_{\nu}=1$, which gives
\begin{equation}
\Delta \nu = |\nu-\nu_0| = \nu_0 \sqrt{2\ln \tau_{\nu_0}} \frac{\sigma_{\rm 1D}}{c}.
\end{equation}
The corresponding range in Doppler shift is
\begin{equation}
\Delta v = \sqrt{2\ln \tau_{\nu_0}} \sigma_{\rm 1D}.
\end{equation}

For this step-function form of $\tau_{\nu}$, the emitted brightness temperature is trivial to compute. At velocity $v$, the brightness temperature is
\begin{equation}
T_{B,v} = \left\{
\begin{array}{ll}
T, \qquad & |v-v_0| < \Delta v \\
0, &  |v-v_0| > \Delta v
\end{array}
\right.
\end{equation}
If we integrate this over all velocities of emitting molecules, we get
\begin{equation}
I_{\rm CO} = \int T_{B,\nu}\, dv = 2 T_B \Delta v = \sqrt{8\ln\tau_{\nu_0}} \sigma_{\rm 1D} T.
\end{equation}
Thus, the velocity-integrated brightness temperature is simply proportional to $\sigma_{\rm 1D}$. The dependence on the line-center optical depth is generally negligible, since that quantity enters only as the square root of the log.

Now let us consider the amount of emission we get per unit column density within our telescope beam. We define this quantity as $X$, and we have
\begin{eqnarray*}
X\mbox{ [cm}^{-2}\mbox{ (K km s}^{-1}\mbox{)}^{-1}\mbox{]} & = & \frac{M/(\mu \pi R^2)}{I_{\rm CO}} \\
& = & 10^5 \frac{(8\ln\tau_{\nu_0})^{-1/2}}{T\mu\pi} \frac{M}{\sigma_{\rm 1D} R^2} \\
& = & 10^5 \frac{(\mu \ln\tau_{\nu_0})^{-1/2}}{T} \sqrt{\frac{5n}{6\pi \avir G}},
\end{eqnarray*}
where $n= 3M/(4\pi R^3)$ is the number density of the cloud, and the factor of $10^5$ comes from the fact that we're measuring $I_{\rm CO}$ in km s$^{-1}$ rather than cm s$^{-1}$. To the extent that all molecular clouds have comparable volume densities on large scales and are virialized, this suggests that there should be a roughly constant CO X factor. If we plug in $T=10$ K, $n=100$ cm$^{-3}$, $\avir=1$, and $\tau_{\nu_0}=100$, this gives $X_{\rm CO}=5\times 10^{19}$ cm$^{-2}$ (K km s$^{-1}$)$^{-1}$.

This is quite a result: it means that we have inferred the mass of a molecular cloud simply by measuring the luminosity it emits in a particular optically thick line. Of course this calculation has a few problems -- we have to assume a volume density, and there are various fudge factors like $a$ floating around.
Moreover, we had to assume virial balance between gravity and internal motions. This implicitly assumes that both surface pressure and magnetic fields are negligible, which they may not be. Making this assumption would necessarily make it impossible to independently check whether molecular clouds are in fact in virial balance between gravity and turbulent motions.

In practice, the way we get around these problems is by determining X factors by empirical calibration. We generally do this by attempting to measure the total gas column density by some tracer that measures all the gas along the line of sight, and then subtracting off the observed atomic gas column -- the rest is assumed to be molecular. One way of doing this is measuring $\gamma$ rays emitted by cosmic rays interacting with the ISM. The $\gamma$ ray emissivity is simply proportional to the number density of hydrogen atoms independent of whether they are in atoms or molecules (since the cosmic ray energy is very large compared to any molecular energy scales). Once produced the $\gamma$ rays travel to Earth without significant attenuation, so the $\gamma$ ray intensity along a line of sight is simply proportional to the total hydrogen column. Another way is to measure the infrared emission from dust grains along the line of sight, which gives the total dust column. This is then converted to a mass column using a dust to gas ratio. Yet a third method is to observe a cloud in multiple molecular lines, some of which are optically thin and some of which are thick, and use the multiple lines in an attempt to determine the absolute mass.

Using any of these techniques in the Milky Way gives $X \approx 2\times 10^{20}$ cm$^{-2}$ (K km s$^{-1}$)$^{-1}$ for the Milky Way, with roughly a factor of 2 scatter on either side depending on the technique used and the assumptions made \cite{abdo10b, blitz07a, draine07a, heyer09a}. These numbers are roughly consistent with our simple model, and the fact that several independent techniques give results that match to a factor of 2 gives us some confidence that the method works.

From this sort of analysis we learn that most of the molecular mass in our galaxy and in similar nearby galaxies is organized into giant clouds with masses of $\sim 10^4 - 10^6$ $\msun$ \cite{rosolowsky05b}.

\section{Physical processes in molecular clouds}

\subsection{Heating and cooling proceses}

The temperature in molecular clouds is set mostly by radiative processes -- adiabatic heating and cooling associated with hydrodynamic motions is generally negligible, as we will show in a moment. Thus we have to consider how clouds can gain and loose heat by radiation. A full treatment of this problem necessarily involves numerical calculations, but we can derive some basic results quite simply.

\subsubsection{Heating by cosmic rays}

In the bulk of the interstellar medium the main source of heating is starlight. However, typical molecular clouds have visual extinctions $A_V \sim 5$, which means that starlight in the interior is reduced to a few percent of the mean interstellar value at visible wavelengths, and to much less than a percent of the interstellar value at the ultraviolet wavelengths that produce most heating. Thus, we can generally neglect starlight as a source of heat (except very near young stars forming within the cloud).

Instead, over the bulk of a molecular cloud's volume, the main source of heating is cosmic rays: relativistic particles accelerated in shocks that are able to penetrate into GMC interiors. How much heat do cosmic rays produce? To answer this question, we must first determine the mechanism by which the gas is heated. The first step in such a heating chain is the interaction of a cosmic ray with an electron, which knocks the electron off a molecule:
\begin{equation}
\mbox{CR}+\mbox{H}_2 \rightarrow \mbox{H}_2^+ + \mbox{e}^- + \mbox{CR}
\end{equation}
The free electron's energy depends only weakly on the CR's energy, and is typically $\sim 30$ eV.

The electron cannot easily transfer its energy to other particles in the gas directly, because its tiny mass guarantees that most collisions are elastic and transfer no energy to the impacted particle. However, the electron also has enough energy to ionize or dissociate other hydrogen molecules, which provides an inelastic reaction that can convert some of its 30 eV to heat. Secondary ionizations do indeed occur, but in this case almost all the energy goes into ionizing the molecule (15.4 eV), and the resulting electron has the same problem as the first one: it cannot effectively transfer energy to the much more massive protons.

Instead, it is secondary dissociations and excitations that wind up being the dominant energy channels. The former reaction is
\begin{equation}
\mbox{e}^- + {\rm H}_2 \rightarrow 2{\rm H} + e^{-}.
\end{equation}
In this reaction any excess energy in the electron beyond what is needed to dissociate the molecule (4.5 eV) goes into kinetic energy of the two recoiling hydrogen atoms, and the atoms, since they are massive, can then efficiently share that energy with the rest of the gas. Alternately, an electron can hit a hydrogen molecule and excite it without dissociating it. The hydrogen molecule then collides with another hydrogen molecule and collisionally de-excites, and the excess energy again goes into recoil, where it is efficiently shared. The reaction is
\begin{eqnarray}
\mbox{e}^- + {\rm H}_2 & \rightarrow & {\rm H}_2^* + e^{-} \\
{\rm H}_2^* + {\rm H}_2 & \rightarrow & 2 {\rm H}_2.
\end{eqnarray}
Summing over all possible transfer channels, and including heating by secondary ionizations too, the energy yield per primary cosmic ray ionization is in the range $7-20$ eV \cite{glassgold73a, dalgarno99a}, depending on the density. These figures are slightly uncertain.

Combining this with the primary ionization rate for cosmic rays in the Milky Way, which is observationally-estimated to be about  $2\times 10^{-17}$ s$^{-1}$ per H nucleus \cite{wolfire10a}, this gives a total heating rate per H nucleus
\begin{equation}
\Gamma_{\rm CR} \sim 10^{-27}\mbox{ erg s}^{-1}.
\end{equation}
The heating rate per unit volume is $\Gamma_{\rm CR} n$, where $n$ is the number density of H nuclei ($=2\times$ the density of H molecules).

\subsubsection{CO cooling}

In molecular clouds there are two main cooling processes: molecular lines and dust radiation. Dust can cool the gas efficiently because dust grains are solids, so they are thermal emitters. However, dust is only able to cool the gas if collisions between dust grains and hydrogen molecules occur often enough to keep them thermally well-coupled. Otherwise the grains cool off, but the gas stays hot. The density at which grains and gas become well-coupled is around $10^4$ cm$^{-3}$ \cite{goldsmith01a}, which is higher than the typical density in a GMC, so we won't consider dust cooling further at this point. We'll return to it in the next section when we discuss collapsing objects, where the densities do get high enough for dust cooling to be important.

The remaining cooling process is line emission, and by far the most important molecule for this purpose is CO, due to its abundance and its ability to radiate even at low temperatures and densities. The physics is fairly simple. CO molecules are excited by inelastic collisions with hydrogen molecules, and such collisions convert kinetic energy to potential energy within the molecule. If the molecule de-excites radiatively, and the resulting photon escapes the cloud, the cloud loses energy and cools.

Let us make a rough attempt to compute the cooling rate via this process. As we mentioned in the last section, a diatomic molecule like CO can be excited rotationally, vibrationally, or electronically. At the low temperatures found in molecular clouds, usually only the rotational levels are important. These are characterized by an angular momentum quantum number $J$, and each level $J$ has a single allowed radiative transition to level $J-1$. Larger $\Delta J$ transitions are strongly suppressed because they require emission of multiple photons to conserve angular momentum.

Unfortunately the CO cooling rate is quite difficult to calculate, because the lower CO lines are all optically thick. A photon emitted from a CO molecule in the $J=1$ state is likely to be absorbed by another one in the $J=0$ state before it escapes the cloud, and if this happens that emission just moves energy around within the cloud and provides no net cooling. The cooling rate is therefore a complicated function of position within the cloud -- near the surface the photons are much more likely to escape, so the cooling rate is much higher than deep in the interior. The velocity dispersion of the cloud also plays a role, since large velocity dispersions Doppler shift the emission over a wider range of frequencies, reducing the probability that any given photon will be resonantly re-absorbed before escaping.

In practice this means that CO cooling rates usually have to be computed numerically, and will depend on the cloud geometry if we want accuracy to better than a factor of $\sim 2$. However, we can get a rough idea of the cooling rate from some general considerations. The high $J$ levels of CO are optically thin, since there are few CO molecules in the $J-1$ states capable of absorbing them, so photons they emit can escape from anywhere within the cloud.
However, the temperatures required to excite these levels are generally high compared to those found in molecular clouds, so there are few molecules in them, and thus the line emission is weak. Moreover, the high $J$ levels also have high critical densities, so they tend to be sub-thermally populated, further weakening the emission.

On other hand, low $J$ levels of CO are the most highly populated, and thus have the highest optical depths. Molecules in these levels produce cooling only if they are within one optical depth the cloud surface. Since this restricts cooling to a small fraction of the cloud volume (typical CO optical depths are many tens for the $1\rightarrow 0$ line), this strongly suppresses cooling.

The net effect of combining the suppression of low $J$ transitions by optical depth effects and of high $J$ transitions by excitation effects is that cooling tends to be dominated a the single line produced by the lowest $J$ level for which the line is not optically thick. This line is marginally optically thin, but is kept close to LTE by the interaction of lower levels with the radiation field. Which line this is depends on the column density and velocity dispersion of the cloud, and detailed calculations show that for typical GMC properties it is generally around $J=5$.

If we assume this dominant cooling level is in LTE, the cooling rate per H nucleus is simply the number of CO molecules per H nucleus times the fraction of molecules in the relevant level, times the emission rate from that level, times the energy per photon:
\begin{equation}
\Lambda_{\rm CO} = x_{\rm CO} (2J+1)\frac{e^{-E_J/(k_B T)}}{Z} A_{\rm J,J-1} (E_{J} - E_{J-1}),
\end{equation}
where $Z$ is the partition function and $x_{\rm CO}$ is the ratio of CO molecules to H nuclei. Note that the factor of $2J+1$ is the degeneracy of level $J$. For a quantum rotator the Einstein $A$'s and energy levels obey
\begin{eqnarray}
E_{J} & = & h B J(J+1) \\
A_{J+1,J} & = & \frac{512\pi^4 B^3 \mu^2}{3 h c^3} \frac{(J+1)^4}{2J+1},
\end{eqnarray}
where $B$ is the rotation constant of the molecule and $\mu$ is its electric dipole moment. For CO, $B=57$ GHz and $\mu=0.112$ Debye.

Plugging these values in, for $J=5\rightarrow 4$ at $T=10$ K we get $\Lambda_{\rm CO} = 1.3\times 10^{-27}$ erg s$^{-1}$ per H nucleus. If we equate the cooling rate to the cosmic ray heating rate of $10^{-27}$ erg s$^{-1}$, which is independent of temperature, we find that heating and cooling balance at $T\approx 10$ K, in good agreement with what we observe. Note that the density does not enter into this, since both $\Gamma$ and $\Lambda_{\rm CO}$ are proportional to density. Thus we expect the equilibrium temperature to be close to density-independent. Due to the exponential dependence, the cooling rate is very temperature-sensitive. If we increase the temperature by a factor of $2$, $\Lambda_{\rm CO}$ rises by a factor of 30, to about $4\times 10^{-26}$ erg s$^{-1}$. Thus it requires a lot of change in heating rate to raise the temperature appreciably.

It is also instructive to consider the timescales implied by these cooling rates. The gas thermal energy per H nucleus is
\begin{equation}
e = \frac{1}{2}\left(\frac{3}{2}k T\right) = 10^{-15} \left(\frac{T}{10\mbox{ K}}\right)\mbox{ erg}
\end{equation}
for a monatomic gas -- and H$_2$ acts like a monatomic gas at low temperature because its rotational degrees of freedom cannot be excited. The factor of $1/2$ comes from 2 H nuclei per H$_2$ molecule. The characteristic cooling time is $t_{\rm cool} = e/\Lambda_{\rm CO}$. Suppose we have gas that is mildly out of equilibrium, say $T=20$ K instead of $T=10$ K. The heating and cooling are far out of balance, so we can ignore heating completely compared to cooling. At the cooling rate of $\Lambda_{\rm CO}=4\times 10^{-26}$ erg s$^{-1}$ for 20 K gas, $t_{\rm cool} = 1.6$ kyr. In contrast, the crossing time for a molecular cloud is $t_{\rm cr} = L/\sigma \sim 7$ Myr for $L=30$ pc and $\sigma = 4$ km s$^{-1}$. The conclusion of this analysis is that radiative effects happen on time scales {\it much} shorter than mechanical ones. Mechanical effects, such as the heating caused by shocks, simply cannot push the gas any significant way out of radiative equilibrium.

\subsection{Flows in Molecular Clouds}

Now that we have satisfied ourselves that the gas in molecular clouds is, for the most part, kept rigidly fixed at a low temperature, let us consider what that implies about the flows of gas in molecular clouds. In the process we will define four important dimensionless numbers that characterize the flow, two each for the magnetic and non-magnetic cases.

\subsubsection{Equations of Motion}

We begin by writing down the basic equations of magnetohydrodynamics that govern flows in molecular clouds. There are three such equations in our case:
\begin{eqnarray}
\frac{\partial\rho}{\partial t} & = & -\nabla \cdot (\rho \vecv) \\
\frac{\partial}{\partial t}(\rho \vecv) & = & -\nabla\cdot(\rho\vecv\vecv) -\nabla P + \frac{1}{4\pi} (\nabla\times\vecB)\times\vecB - \rho \nabla \phi + \rho \nu \nabla^2\vecv \\
\frac{\partial \vecB}{\partial t} & = & -\nabla\times(\vecB\times\vecv) - \nabla\times(\bm{\eta} : \nabla\times\vecB) \\
\nabla^2 \phi & = & -4\pi G \rho.
\end{eqnarray}
The quantities here are the density $\rho$, the velocity $\vecv$, the pressure $P$, the magnetic field $\vecB$, the gravitational potential $\phi$, the kinematic viscosity $\nu$, and the magnetic resistivity $\bm{\eta}$. Note that, in general, $\bm{\eta}$ can be a tensor, and the colon represents tensor contraction. Since the temperature is fixed by radiative effects, the equation of state is simple. We characterize the temperature by a sound speed $c_s$, which is related to the pressure by
\begin{equation}
P = \rho c_s^2.
\end{equation}

Physically, the first equation represents conservation of matter. It states that the rate of change in density at a given point, $\partial \rho/\partial t$, is equal to the rate at which mass flows toward or away from the point, $-\nabla \cdot(\rho\vecv)$. Similarly, the second equation represents conservation of momentum. It states that the rate of change of the momentum is equal to the rate at which momentum is advected away by the flow, plus four remaining terms on the right hand side, which represent pressure forces, Lorentz (magnetic) forces, gravitational forces, and viscous forces. Finally, the third equation is the induction equation, and it states that the time rate of change of the magnetic field is equal to the rate at which the field is carried along by the fluid plus the rate at which the field is either dissipated or diffused by resistance in the fluid. Finally, the last equation gives the gravitational potential due to the matter in the cloud.

Solving these four equations in general is not feasible, except by numerical simulation. Instead, we will simply analyze them to try to derive some general results and gain some insight.

\subsubsection{Dimensionless Numbers}

We start by considering which terms are important. Suppose our system is characterized by a size scale $L$, a velocity scale $V$, and a magnetic field strength $B$. In a molecular cloud, we might have $L\sim 30$ pc, $V\sim 3$ km s$^{-1}$, and $B\sim 10$ $\mu$G. The natural scale for spatial derivatives is $1/L$, so to figure out how big terms are to the order of magnitude level, we can take on of our equations and replace all the spatial derivatives with $1/L$. Doing so with the momentum equation, and dropping the gravitational force term for now, the terms on the right hand side are
\begin{equation}
\rho\frac{V^2}{L} + \rho \frac{c_s^2}{L} + \frac{B^2}{L} + \rho \nu \frac{V}{L^2}.
\end{equation}
These terms represent, from left to right: advection of momentum by fluid flows, changes in momentum due to pressure forces, changes in momentum due to magnetic forces, and changes in momentum due to viscous forces.

We can get a learn a lot simply by figuring out which of these terms are important and which are not. Only terms that are important will contribute to the time derivative, and thus to the evolution of the system. Let's start by comparing the first and second terms. Clearly the ratio of the first term, representing advection, to the second term, pressure, is of order $(V/c_s)^2$. We define the square root of this ratio as the {\bf Mach Number} of the flow:
\begin{equation}
\mathcal{M} = \frac{V}{c_s}.
\end{equation}
The significance of $\mathcal{M}$ is that, when $\mathcal{M} \gg 1$, the advection term, representing momentum changing at a given position because of fluid motions, is much more important than the pressure term, representing changes in momentum at a given point due to pressure forces. Thus when $\mathcal{M}$ pressure becomes unimportant. We have already seen that the sound speed in a molecular cloud is $c_s \approx 0.2$ km s$^{-1}$, so $\mathcal{M} \sim 10$, and we conclude that the pressure term is sub-dominant by a factor of $\mathcal{M}^2 \sim 100$. Thus we reach our first interesting conclusion based on dimensionless numbers: molecular cloud flows are highly supersonic, and this means that pressure forces are unimportant.

\begin{figure}
\includegraphics[height=.3\textheight]{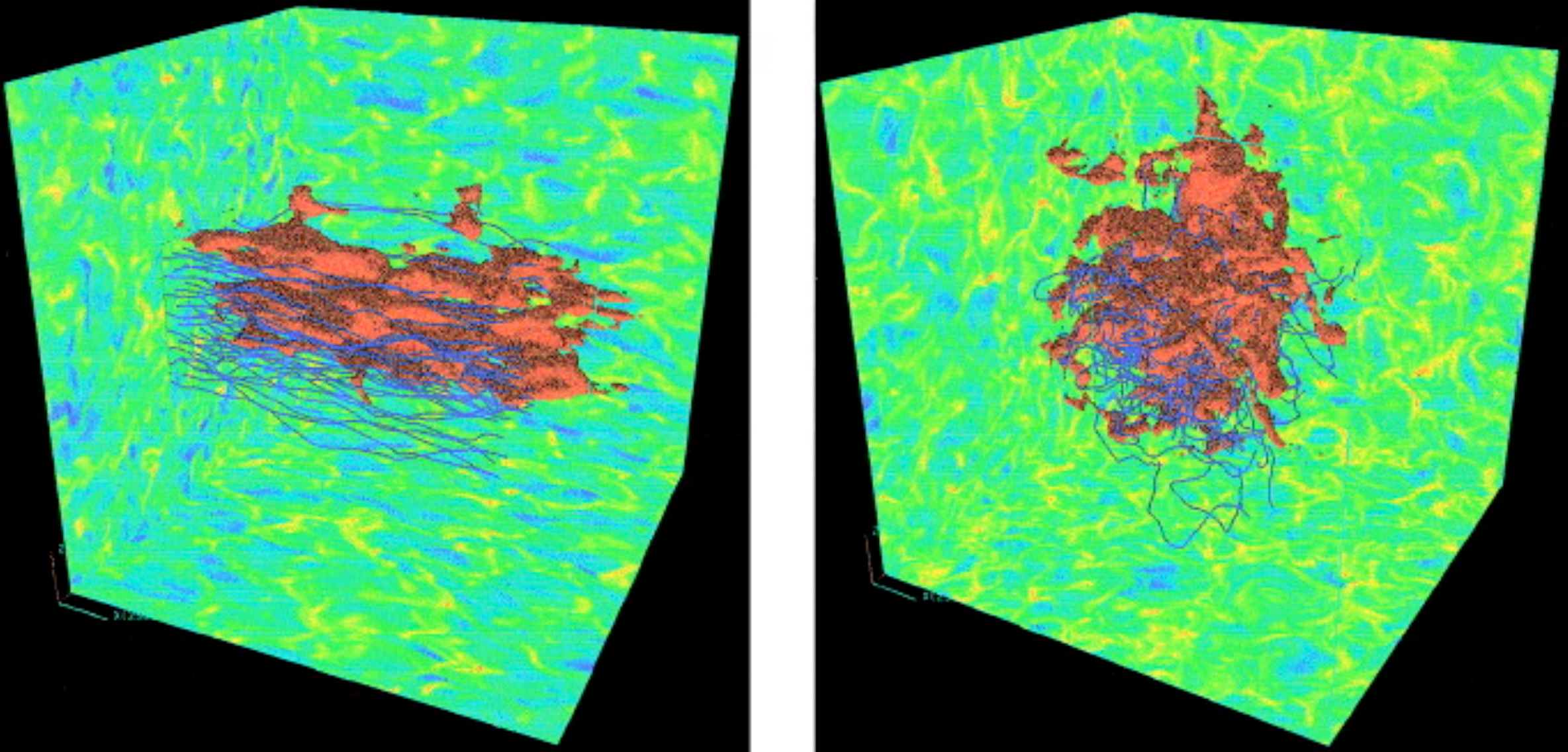}
  \caption{Two simulations of driven magnetohydrodynamic turbulence, one with a low Alfv\'en Mach number (left) and one with a high Alfv\'en Mach number (right). The colored box walls show the logarithm of density, The blue lines are magnetic field lines. The red surfaces show the distribution of a passive contaminant that has been added to the flow. Reprinted with permission from the AAS from \cite{stone98a}.
  \label{fig:alfvenmach}}
\end{figure}

Now that we have determined pressure forces are unimportant, let us consider magnetic forces. Clearly the ratio of those two terms is of order $B^2/(\rho V^2)$. We use the square root of this ratio to define the {\bf Alfv\'en Mach Number}, after Hannes Alfv\'en, the father of magnetohydrodynamics. Formally,
\begin{equation}
\mathcal{M}_A = \frac{V}{v_A},
\end{equation}
where $v_A = B/\sqrt{4\pi \rho}$ is called the Alfv\'en speed, which plays a role analogous to the sound speed in non-magnetized fluid dynamics. If $\mathcal{M}_A \gg 1$, this means that the advection term is much more important than the magnetic term, so magnetic forces are unimportant. On the other hand $\mathcal{M}_A \ltsim 1$ means that magnetic forces are important, and this will tend to force the flow to follow magnetic field lines. The resulting flow morphology will be quite different, as shown in Figure \ref{fig:alfvenmach}. Evaluating our terms for a molecular cloud, we have $v_A = B/\sqrt{4\pi \rho} \sim 2$ km s$^{-1}$, so $\mathcal{M}_A\sim 1$. Thus we conclude that magnetic forces are not negligible in molecular clouds. They are comparable in importance to advection.

Third, let us compare the advection term to the viscous term. Taking the ratio of these two gives
\begin{equation}
\mbox{Re} = \frac{\mbox{advection}}{\mbox{viscous forces}} = \frac{LV}{\nu}.
\end{equation}
The quantity we have defined is called the {\bf Reynolds Number}, and it clearly characterizes the importance of viscous forces. If $\mbox{Re}\gg 1$, this means that viscosity is unable to change fluid velocities on a timescale comparable to the natural crossing timescale of the flow. We can also think of the Reynolds number as describing a characteristic size scale $L\sim \nu/V$ on which motions are damped. Motions larger than this scale are unimpeded by viscosity, while smaller scale motions are damped out.

For diffuse gases, the kinematic viscosity $\nu = 2\overline{u}\lambda$, where $\overline{u}$ is the RMS particle speed and $\lambda$ is the mean free path. The former is comparable to the sound speed, $\overline{u}\sim 0.2$ km s$^{-1}$, while the latter is of order the inverse of the cross section times the particle density, $\lambda\sim 1/(n\sigma)$. For a typical molecule size of 1 nm and a density of 100 cm$^{-2}$, this gives $\lambda \sim 10^{12}$ cm. Thus we have $\nu\sim 10^{16}$ cm$^2$ s$^{-1}$, and putting this together with our characteristic size and velocity scales gives $\mbox{Re} \sim 10^9$. We therefore learn that molecular cloud flows are extremely non-viscous. An important implication of this is that they are almost certainly turbulent, since essentially all flows with $\mbox{Re} \gtsim 10^3$ are observed to be turbulent.

Finally, let's apply our non-dimensionalization treatment to the induction equation. Doing so, the two terms on the right hand side are of order
\begin{equation}
\frac{BV}{L} + \eta \frac{B}{L^2}.
\end{equation}
In analogy with the Reynolds number, we define the ratio of these two terms by the {\bf Magnetic Reynolds Number}
\begin{equation}
\mbox{Rm} = \frac{LV}{\eta}.
\end{equation}
The significance of Rm is that, when it is large, the advection term in the induction equation is much larger than the resistive one, and this means that magnetic field is simply carried around by the fluid. We describe this situation as flux-freezing, meaning that lines of magnetic flux passing through a fluid element move with that fluid element at all times.

The value of the magnetic Reynolds number depends on the resistivity, which is tricky to calculate. We won't do so in these lectures, but we can outline the main process that contributes to it: ion-neutral drift, also known as ambipolar diffusion. In a molecular cloud, most of the gas is actually neutral, not ionized, and so it doesn't feel magnetic forces. Only the ions do. However, the ions collide with the neutrals, and if those collisions are frequent enough then they will transmit the magnetic force to the neutrals. However, this process isn't perfect. If there are too few ions, then neutrals may go a long time before encountering an ion, and they will begin to drift with respect to the ions, since they ions are being pulled by magnetic forces that the neutrals don't feel. For this reason the resistivity depends on the ionization fraction, with lower ionization fractions giving higher resistivities.

In a molecular cloud the ionization fraction is controlled by the balance between cosmic ray ionizations and recombinations of electrons with ions, and calculations of this process (e.g.\ \cite{tielens05a}) suggest that, at a density of $n\sim 100$ cm$^{-3}$, typical ionization fractions are about $10^{-6}$. That might not seem like much, but it's enough to produce a fairly small resistivity -- working through the calculation gives ${\rm Rm}\sim 50$ for our typical parameters. We therefore conclude that resistivity is not significant on GMC scales, and flux-freezing holds. However, note that, as with viscosity, we can define a characteristic scale where resistivity does become important, which is $L/{\rm Rm}$. This is $\sim 0.5$ pc, and this means that, on small scales, we expect magnetic fields to begin to decouple from the gas. We'll return to this issue later on.

\subsection{The Virial Theorem}

As our final topic in this lecture, we will derive a theorem that describes the large-scale behavior of molecular clouds. This is the virial theorem, which is a sort of integrated version of the equations of motion. Like the equations of motion, there is both an Eulerian form and a Lagrangian form of the virial theorem, depending on which version of the equations of motion we start with. We'll derive the Eulerian form here, but the derivation of the Lagrangian form proceeds in a similar manner, and can be found in many standard textbooks. This derivation follows that of McKee \& Zweibel \cite{mckee92a}.

To derive the virial theorem, we begin with the MHD equations of motion, without either viscosity or resistivity (since neither of these are important for GMCs on large scales) but with gravity. We leave in the pressure forces, even though they are small, because they're also trivial to include. Thus we have
\begin{eqnarray}
\frac{\partial\rho}{\partial t} & = & -\nabla \cdot (\rho \vecv) \\
\frac{\partial}{\partial t}(\rho \vecv) & = & -\nabla\cdot(\rho\vecv\vecv) -\nabla P + \frac{1}{4\pi} (\nabla\times\vecB)\times\vecB - \rho \nabla \phi.
\end{eqnarray}
Here $\phi$ is the gravitational potential, so $-\rho \nabla \phi$ is the gravitational force per unit volume.

These equations are the Eulerian equations written in conservative form. The first is conservation of mass: it says that the rate of change of density is equal to the flux of mass into or out of a given volume. The second is conservation of momentum: it says that the rate of change of momentum is given by the flux of momentum in or out of a given volume plus the changes in momentum due to gas pressure, magnetic forces, and gravitational forces. Note here that in the second equation the term $\vecv\vecv$ is a tensor -- in tensor notation, its elements are $v_i v_j$.

Before we begin, life will be a bit easier if we re-write the entire second equation in a manifestly tensorial form -- this simplifies the analysis tremendously. To do so, we define two tensors: the fluid pressure tensor $\vecPi$ and the Maxwell stress tensor $\vecT_M$, as follows:
\begin{eqnarray}
\vecPi & \equiv & \rho \vecv\vecv + P\vecI \\
\vecT_M & \equiv & \frac{1}{4\pi} \left(\vecB\vecB - \frac{B^2}{2}\vecI\right)
\end{eqnarray}
Here $\vecI$ is the identity tensor. In tensor notation, these are
\begin{eqnarray}
(\vecPi)_{ij} & \equiv & \rho v_i v_j + P \delta_{ij} \\
(\vecT_M)_{ij} & \equiv & \frac{1}{4\pi} \left(B_i B_j - \frac{1}{2}B_k B_k \delta_{ij}\right)
\end{eqnarray}
With these definitions, the momentum equation just becomes
\begin{equation}
\frac{\partial}{\partial t}(\rho \vecv) = -\nabla\cdot(\vecPi-\vecT) - \rho \nabla\phi.
\end{equation}

The substitution for $\vecPi$ is obvious. The equivalence of $\nabla\cdot\vecT_M$ to $1/(4\pi) (\nabla\times\vecB)\times\vecB$ is easy to establish with a little vector manipulation, which is most easily done in tensor notation:
\begin{eqnarray}
(\nabla\times\vecB)\times\vecB & = & \epsilon_{ijk} \epsilon_{jmn} \left(\frac{\partial}{\partial x_m}B_n\right) B_k \\
& = & -\epsilon_{ijk} \epsilon_{jmn} \left(\frac{\partial}{\partial x_m}B_n\right) B_k \\
& = & (\delta_{in}\delta_{km}-\delta_{im}\delta_{kn})\left(\frac{\partial}{\partial x_m}B_n\right) B_k \\
& = & B_k\frac{\partial}{\partial x_k} B_i - B_k\frac{\partial}{\partial x_i} B_k \\
& = & \left(B_k\frac{\partial}{\partial x_k} B_i + B_i \frac{\partial}{\partial x_k} B_k\right) - B_k\frac{\partial}{\partial x_i} B_k \\
& = & \frac{\partial}{\partial x_k}\left(B_i B_k\right) -\frac{1}{2} \frac{\partial}{\partial x_i} \left(B_k^2\right)\\
& = & \nabla\cdot \left(\vecB\vecB - \frac{B^2}{2}\right)
\end{eqnarray}

To derive the virial theorem, we begin by imagining a cloud of gas enclosed by some fixed volume $V$. The surface of this volume is $S$. We want to know how the overall distribution of mass changes within this volume, so we begin by writing down a quantity the represents the mass distribution. This is the moment of inertia:
\begin{equation}
I = \int_V \rho r^2\, dV.
\end{equation}

We want to know how this changes in time, so we take its time derivative:
\begin{eqnarray}
\dot{I} & = & \int_V \frac{\partial\rho}{\partial t} r^2 \,dV \\
& = & -\int_V \nabla \cdot (\rho \vecv) r^2\, dV \\
& = & -\int_V \nabla \cdot (\rho \vecv r^2)\, dV + 2\int_V \rho \vecv\cdot \vecr\, dV \\
& = & -\int_S (\rho \vecv r^2)\cdot d\vecS + 2\int_V \rho \vecv\cdot \vecr\, dV
\end{eqnarray}
In the first step we used the fact that the volume $V$ does not vary in time to move the time derivative inside the integral. Then in the second step we used the equation of mass conservation to substitute. In the third step we brought the $r^2$ term inside the divergence. Finally in the fourth step we used the divergence theorem to replace the volume integral with a surface integral.

Now we take the time derivative again, and multiply by $1/2$ for future convenience:
\begin{eqnarray}
\frac{1}{2}\ddot{I} & = & -\frac{1}{2} \int_S r^2 \frac{\partial}{\partial t}(\rho\vecv)\cdot d\vecS +
\int_V \frac{\partial}{\partial t}(\rho\vecv)\cdot\vecr \, dV \\
& = & -\frac{1}{2} \frac{d}{dt} \int_S r^2 (\rho\vecv)\cdot d\vecS -
\int_V \vecr \cdot \left[\nabla\cdot(\vecPi-\vecT_M)+ \rho\nabla \phi \right] \, dV
\end{eqnarray}

The term involving the tensors is easy to simplify using a handy identity, which applies to an arbitrary tensor. This is a bit easier to follow in tensor notation:
\begin{eqnarray}
\int_V \vecr\cdot \nabla\cdot \vecT \, dV & = & \int_V x_i \frac{\partial}{\partial x_j} T_{ij}\,dV \\
& = & \int_V \frac{\partial}{\partial x_j}(x_i T_{ij})\,dV - \int_V T_{ij} \frac{\partial}{\partial x_j}x_i \, dV \\
& = & \int_S x_i T_{ij} \,dS_j - \int_V \delta_{ij} T_{ij} \, dV \\
& = & \int_S \vecr\cdot\vecT\cdot d\vecS - \int_V \mbox{Tr}\; \vecT \, dV,
\end{eqnarray}
where $\mbox{Tr}\;\vecT = T_{ii}$ is the trace of the tensor $\vecT$.

Applying this to our result our tensors, we note that
\begin{eqnarray}
\mbox{Tr}\; \vecPi & = & 3P + \rho v^2 \\
\mbox{Tr}\; \vecT_M & = & -\frac{B^2}{8\pi}
\end{eqnarray}
Inserting this result into our expression for $\ddot{I}$ give the virial theorem, which I will write in a more suggestive form to make its physical interpretation clearer:
\begin{equation}
\label{eq:virialthm}
\frac{1}{2}\ddot{I} = 2(\mathcal{T} - \mathcal{T}_S) + \mathcal{M} + \mathcal{W} - \frac{1}{2}\frac{d}{dt} \int_S (\rho\vecv r^2)\cdot d\vecS,
\end{equation}
where
\begin{eqnarray}
\mathcal{T} & = & \int_V\left(\frac{1}{2}\rho v^2 + \frac{3}{2} P\right)\, dV \\
\mathcal{T}_S & = & \int_S \vecr \cdot \vecPi \cdot d\vecS\\
\mathcal{M} & = & \frac{1}{8\pi} \int_V B^2 \,dV + \int_S \vecr\cdot \vecT_M\cdot d\vecS\\
\mathcal{W} & = & -\int_V \rho \vecr\cdot\nabla\phi\,dV
\end{eqnarray}

Written this way, we can give a clear interpretation to what these terms mean. $\mathcal{T}$ is just the total kinetic plus thermal energy of the cloud. $\mathcal{T}_S$ is the confining pressure on the cloud surface, including both the thermal pressure and the ram pressure of any gas flowing across the surface. $\mathcal{M}$ is the the difference between the magnetic pressure in the cloud interior, which tries to hold it up, and the magnetic pressure plus magnetic tension at the cloud surface, which try to crush it. $\mathcal{W}$ is the gravitational energy of the cloud. If there is no external gravitational field, and $\phi$ comes solely from self-gravity, then $\mathcal{W}$ is just the gravitational binding energy. The final integral represents the rate of change of the momentum flux across the cloud surface.

$\ddot{I}$ is the integrated form of the acceleration. For a cloud of fixed shape, it tells us the rate of change of the cloud's expansion of contraction. If it is negative, the terms that are trying to collapse the cloud (the surface pressure, magnetic pressure and tension at the surface, and gravity) are larger, and the cloud accelerates inward. If it is positive, the terms that favor expansion (thermal pressure, ram pressure, and magnetic pressure) are larger, and the cloud accelerates outward. If it is zero, the cloud neither accelerates nor decelerates.

We get a particularly simple form of the virial theorem if there is no gas crossing the cloud surface (so $\vecv=0$ at $S$) and if the magnetic field at the surface to be a uniform value $B_0$. In this case the virial theorem reduces to
\begin{equation}
\frac{1}{2}\ddot{I} = 2(\mathcal{T} - \mathcal{T}_S) + \mathcal{M} + \mathcal{W}
\end{equation}
with
\begin{eqnarray}
\mathcal{T}_S & = & \int_S rP dS\\
\mathcal{M} & = & \frac{1}{8\pi} \int_V (B^2-B_0^2) \,dV.
\end{eqnarray}
In this case $\mathcal{T}_S$ just represents the mean radius times pressure at the virial surface, and $\mathcal{M}$ just represents the total magnetic energy of the cloud minus the magnetic energy of the background field over the same volume.

Notice that, if a cloud is in equilibrium ($\ddot{I}=0$) and magnetic and surface forces are negligible, then we have $2\mathcal{T} = -\mathcal{W}$, which is what went into our definition of the virial ratio above: $\alpha_{\rm vir} = 2\mathcal{T}/|\mathcal{W}|$.

\section{Molecular Cloud Collapse}

We are now at the point where we can discuss why molecular clouds collapse to form stars, and explore the basic physics of that collapse. We will first look at instabilities that cause collapse, and then discuss what happens when collapse occurs.

\subsection{Stability Conditions}

In considering whether molecular clouds can collapse, it is helpful to look at the virial theorem, equation (\ref{eq:virialthm}).
We can group the terms on the right hand side into those that are generally or always positive, and thus oppose collapse, and those that are generally or always negative, and thus encourage it. The main terms opposing collapse are $\mathcal{T}$, which contains parts describing both thermal pressure and turbulent motion, and $\mathcal{M}$, which describes magnetic pressure and tension. The main terms favoring collapse are $\mathcal{W}$, representing self-gravity, and $\mathcal{T}_S$, representing surface pressure. The final term, the surface one, could be positive or negative depending on whether mass is flowing into our out of the virial volume. We will begin by examining the balance among these terms, and the forces they represent.

\subsubsection{Thermal Pressure: the Bonnor-Ebert Mass}

To begin with, consider a cloud where magnetic forces are negligible, so we need only consider pressure and gravity. For simplicity we'll adopt a spherical geometry, since more complex geometries only change the result by factors of order unity, and we will neglect the flux of mass across the cloud surface, since on average that contributes neither to support nor to collapse. Thus we have a spherical cloud of mass $M$ and radius $R$, bounded by an external medium that exerts a pressure $P_s$ at its surface. The material in the cloud has a one-dimensional velocity dispersion $\sigma$ (including thermal and non-thermal motions). With this assumption, the terms that appear on the right-hand side of the virial theorem are
\begin{eqnarray}
\mathcal{T} & = & \int_V \left(\frac{1}{2} \rho v^2 + \frac{3}{2} P\right) = \frac{3}{2} M \sigma^2 \\
\mathcal{T}_S & = & \int_S \vecr \cdot \vecPi \cdot d\vecS = 4\pi R^3 P_S \\
\mathcal{W} & = & -\int_V \rho \vecr\cdot\nabla\phi \, dV = -a\frac{G M^2}{R},
\end{eqnarray}
where $a$ is a constant of order unity that depends on the internal density distribution of the cloud.

If we wish the cloud to be in virial equilibrium, then we have
\begin{equation}
0 = \frac{3}{2} M \sigma^2 - 4\pi R^3 P_S -a\frac{G M^2}{R},
\end{equation}
which we can re-arrange to
\begin{equation}
P_S = \frac{1}{4\pi R^3} \left(\frac{3}{2}M\sigma^2 - a \frac{GM^2}{R}\right)
\end{equation}
This expression has an interesting feature. If we consider a cloud of fixed $M$ and $\sigma$ and vary the radius $R$, we find that $P_S$ has a maximum value 
\begin{equation}
P_S = \frac{3^7 \sigma^8}{2^{14} \pi a^3 G^3 M^2}.
\end{equation}
We can understand what is going on physically as follows. Consider starting a cloud at very large radius $R$. In this case its self-gravity is negligible, so the second term in parentheses is can be dropped, but the mean density is very low and so the pressure is low. As we decrease the radius the pressure rises initially, but as the radius gets larger self-gravity starts to become important, and more and more of the cloud's internal pressure goes to holding it up against self-gravity, rather than against the external surface pressure. Eventually we reach a point where further contraction is counter-productive and actually lowers the surface pressure.

Now turn this around: if we consider a cloud with a fixed mass and internal velocity dispersion, and we vary the surface pressure, this means that, once the pressure exceeds a fixed value, there is no way that the cloud can remain in virial equilibrium. Instead, it must collapse. The maximum possible pressure is a decreasing function of mass, so larger and larger masses become progressively more and more unstable.

In order to be more quantitative about this, we need to know the value of $a$, which depends on the internal density distribution. We can solve for this by writing down the equation of hydrostatic balance for the cloud and finding the value of $a$ from the self-consistently determined density distribution. We won't do so in these notes, but the result can be found in standard textbooks. The result is that the maximum mass that can be held up in an environment where the surface pressure is $P_S$ is
\begin{equation}
M_{\rm BE} \approx 1.18 \frac{\sigma^4}{G^{3/2} P_S^{1/2}} = 0.47\msun \left(\frac{\sigma}{0.2\mbox{ km s}^{-1}}\right)^4 \left(\frac{P_S/k_B}{10^6\mbox{ K cm}^{-3}}\right)^{-1/2}.
\end{equation}
This is known as the Bonnor-Ebert mass. The scalings chosen for $\sigma$ and $P_S$ are typical of the thermal sound speed and the pressure in molecular clouds, and it is certainly interesting that when we plug in these values we get something like the typical mass of a star.

Of course if we put in the typical observed velocity dispersion in molecular clouds, $\sigma \sim$ a few km s$^{-1}$, we get a vastly larger mass -- more like $10^4 - 10^6$ $\msun$, the mass of a GMC. This makes sense. It's equivalent to our statement from above that the virial ratios of molecular clouds are about unity. However,  turbulent support is a tricky thing. It doesn't work everywhere. In some places the turbulent flows come together and cancel out, and in those places the velocity dispersion drops to the thermal value, and collapse can occur if the mass exceeds the Bonnor-Ebert mass. We'll return to this idea of large-scale support by turbulence coupled with localized collapse in the final section.

\subsubsection{Magnetic Support: the Magnetic Critical Mass}

Now let us consider a cloud where the magnetic term in the virial theorem greatly exceeds the kinetic one. Again, we'll consider a simple case to get the basic scalings: a uniform spherical cloud of radius $R$ threaded by a magnetic field $\vecB$. We imagine that $\vecB$ is uniform inside the cloud, but that outside the cloud the field lines quickly spread out, so that the magnetic field drops down to some background strength $\vecB_0$, which is also uniform but has a magnitude much smaller than $\vecB$.

The magnetic term in the virial theorem is
\begin{equation}
\mathcal{M} = \frac{1}{8\pi} \int_V B^2 \,dV + \int_S \vecx \cdot \vecT_M \cdot d\vecS
\end{equation}
where
\begin{equation}
\vecT_M = \frac{1}{4\pi} \left(\vecB\vecB - \frac{B^2}{2}\vecI\right).
\end{equation}
If the field inside the cloud is much larger than the field outside it, then the first term, representing the integral of the magnetic pressure within the cloud, is
\begin{equation}
\frac{1}{8\pi} \int_V B^2\, dV \approx \frac{B^2 R^3}{6}.
\end{equation}
Here we have dropped any contribution from the field outside the cloud. The second term, representing the surface magnetic pressure and tension, is
\begin{equation}
\int_S \vecx \cdot \vecT_M \cdot d\vecS = \int_S \frac{B_0^2}{8\pi} \vecx \cdot d\vecS
\approx \frac{B_0^2 R_0^3}{6}.
\end{equation}

Since the field lines that pass through the cloud must also pass through the virial surface, it is convenient to rewrite everything in terms of the magnetic flux. The flux passing through the cloud is $\Phi_B = \pi B R^2$, and since these field lines must also pass through the virial surface, we must have $\Phi_B = \pi B_0 R_0^2$ as well. Thus, we can rewrite the magnetic term in the virial theorem as
\begin{equation}
\mathcal{M} \approx \frac{B^2 R^3}{6} - \frac{B_0^2 R_0^2}{6} = \frac{1}{6\pi^2} \left(\frac{\Phi_B^2}{R} - \frac{\Phi_B^2}{R_0}\right) \approx \frac{\Phi_B^2}{6\pi^2 R}.
\end{equation}
In the last step we used the fact that $R \ll R_0$ to drop the $1/R_0$ term.

Now let us compare this to the gravitational term, which is
\begin{equation}
\mathcal{W} = -\frac{3}{5} \frac{GM^2}{R}
\end{equation}
for a uniform cloud of mass $M$. Comparing these two terms, we find that
\begin{equation}
\mathcal{M}+\mathcal{W} = \frac{\Phi_B^2}{6\pi^2 R} - \frac{3}{5} \frac{GM^2}{R} \equiv \frac{3}{5}\frac{G}{R} \left(M_{\Phi}^2-M^2\right)
\end{equation}
where
\begin{equation}
M_{\Phi} \equiv \sqrt{\frac{5}{2}} \left(\frac{\Phi_B}{3 \pi G^{1/2}}\right)
\end{equation}
We call $M_{\Phi}$ the magnetic critical mass.

Since both $\Phi_B$ does not change as a cloud expands or contracts (due to flux-freezing), this magnetic critical mass does not change either. The implication of this is that clouds that have $M>M_{\Phi}$ always have $\mathcal{M}+\mathcal{W} < 0$. The magnetic force is unable to halt collapse no matter what. Clouds that satisfy this condition are called magnetically supercritical, because they are above the magnetic critical mass $M_{\Phi}$. Conversely, if $M<M_{\Phi}$, then $\mathcal{M}+\mathcal{W} > 0$, and gravity is weaker than magnetism.

Clouds satisfying this condition are called subcritical. For a subcritical cloud, since $\mathcal{M}+\mathcal{W} \propto 1/R$, this term will get larger and larger as the cloud shrinks. In other words, not only is the magnetic force resisting collapse is stronger than gravity, it becomes larger and larger without limit as the cloud is compressed to a smaller radius. Unless the external pressure is also able to increase without limit, which is unphysical, then there is no way to make a magnetically subcritical cloud collapse. It will always stabilize at some finite radius. The only way to get around this is to change the magnetic critical mass, which requires changing the magnetic flux through the cloud. This is possible only via ambipolar diffusion or some other non-ideal MHD effect that violates flux-freezing.

Of course our calculation is for a somewhat artificial configuration of a spherical cloud with a uniform magnetic field. In reality a magnetically-supported cloud will not be spherical, since the field only supports it in some directions, and the field will not be uniform, since gravity will always bend it some amount. Figuring out the magnetic critical mass in that case requires solving for the cloud structure numerically. A calculation of this effect by Tomisaka et al. \cite{tomisaka98a} gives
\begin{equation}
M_{\Phi} = 0.12\frac{\Phi_B}{G^{1/2}}
\end{equation}
for clouds for which pressure support is negligible. The numerical coefficient we got for the uniform cloud case is $0.17$, so this is obviously a small correction. It is also possible to derive a combined critical mass that incorporates both the flux and the sound speed, and which limits to the Bonnor-Ebert mass for negligible field and the magnetic critical mass for negligible pressure.

\begin{figure}
\includegraphics[height=.3\textheight]{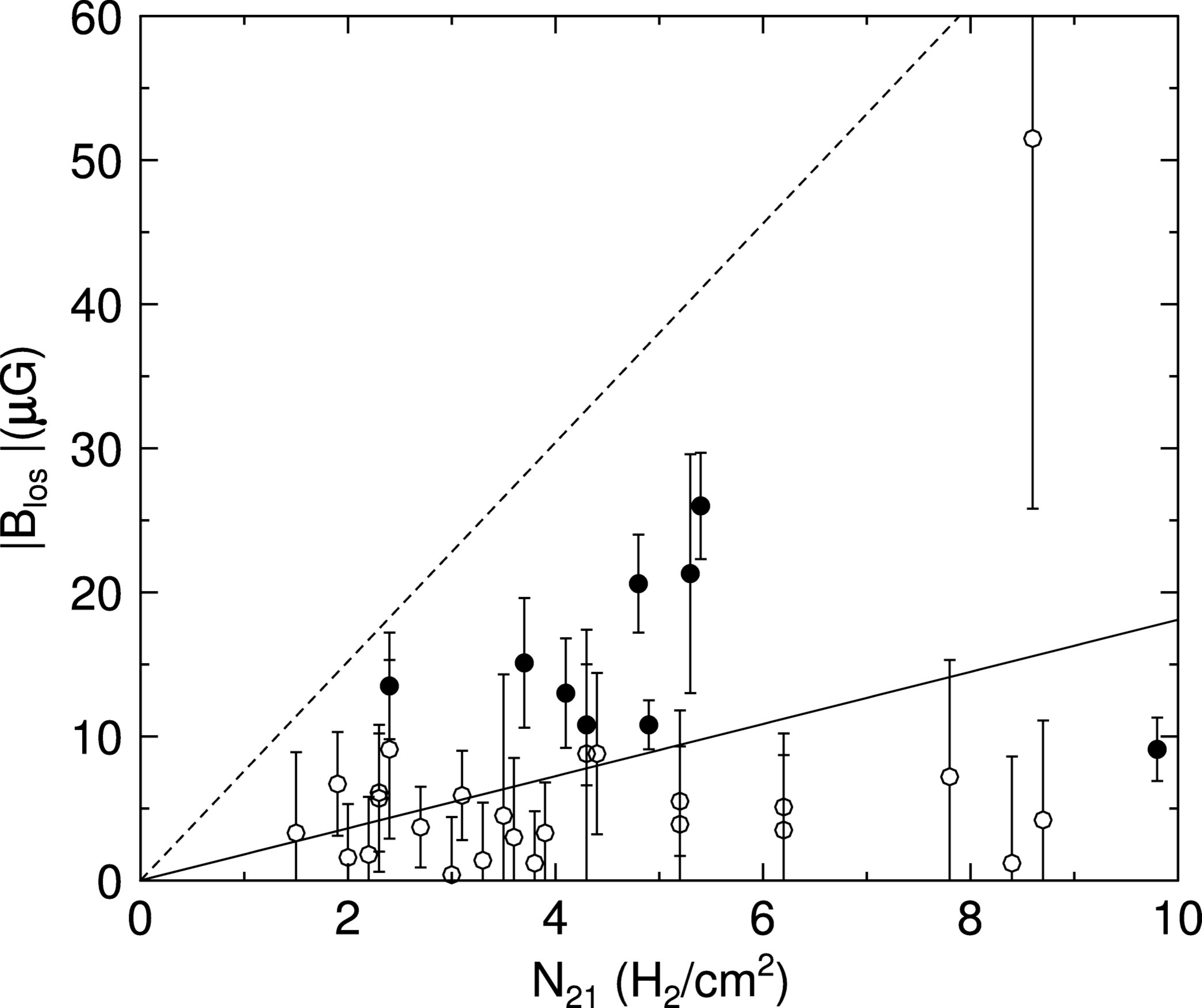}
  \caption{Measurements of the line of sight magnetic field strength $B_{\rm los}$ in a sample of molecular cloud cores, as a function of the number density of H$_2$ molecules $N_{21}$ (in units of $10^{21}$ molecules cm$^{-2}$). Filled points show detections, while empty points show upper limits. The dashed line is the division between magnetically subcritical (above the line) and magnetically supercritical (below the line). Error bars are 1$\sigma$. The solid line is a linear fit to the data. The magnetic field strengths shown here were measured using the Zeeman effect. Reprinted with permission from the AAS from \cite{troland08a}.
  \label{fig:zeeman}}
\end{figure}

Given that a sufficiently strong magnetic field can prevent the collapse of a cloud, it is a critical question whether molecular clouds are super- or subcritical. This must be answered empirically. Observations of magnetic fields in molecular clouds are extremely difficult, and we will not take the time to go into the various techniques that are used. Nonetheless, the observations at this point do seem to show that molecular clouds are magnetically supercritical, although not by a lot -- see Figure \ref{fig:zeeman}. However, since this is a difficult observation, this interpretation of the data is not universally accepted.

\subsection{Collapsing Cores}

\subsubsection{Spherical collapse}

\begin{figure}
\includegraphics[height=.3\textheight]{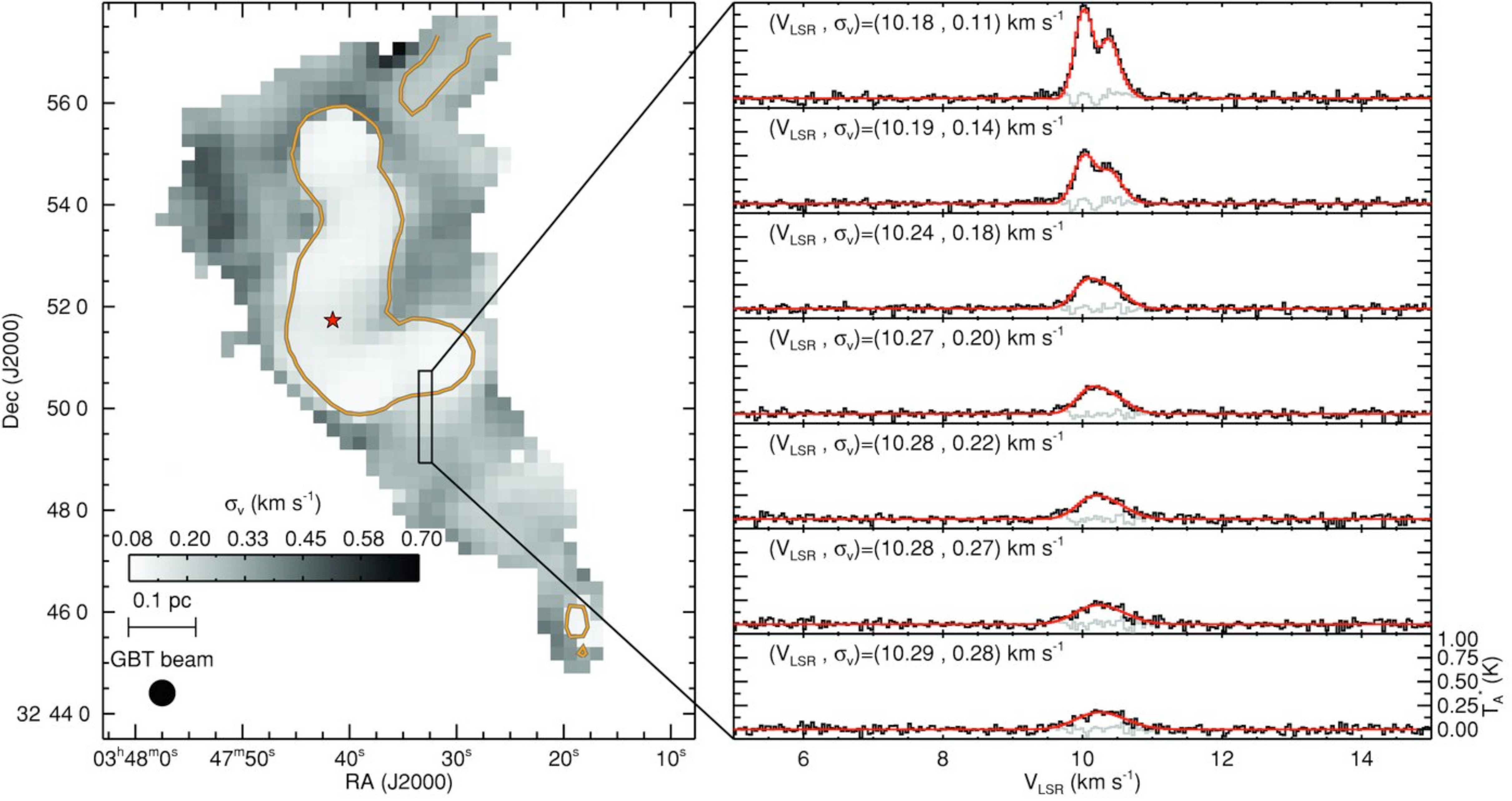}
  \caption{Measurements of the velocity dispersion of the gas in and around a dense core, measured using ammonia (NH$_3$) lines. The grayscale image on the left shows the velocity dispersion as a function of position, and on the right the figure shows the measured spectra at the indicated positions. Reprinted with permission from the AAS from \cite{pineda10a}.
  \label{fig:core}}
\end{figure}

The simplest case to think about, and a good one to understand some of the basic physical processes, is the collapse of a non-rotating, non-turbulent, isothermal spherical core without a magnetic field, supported by thermal pressure. Of course none of these assumptions are strictly true, but they give us a place to begin our study. Moreover, the assumption that collapsing regions, called cores, are not strongly supersonic is reasonable, since collapse tends to occur in places where the turbulent velocities cancel. Observations show this, e.g.\ as illustrated in Figure \ref{fig:core}.

\paragraph{Density and Velocity Profiles}

Consider a sphere of gas with an initial density distribution $\rho(r)$. We would like to know how the gas moves under the influence of gravity and thermal pressure, under the assumption of spherical symmetry. For convenience we define the enclosed mass
\begin{equation}
M_r =  \int_0^r 4\pi r'^2 \rho(r') \, dr'
\end{equation}
or equivalently
\begin{equation}
\frac{\partial M_r}{\partial r} = 4\pi r^2 \rho.
\end{equation}
The equation of mass conservation for the gas in spherical coordinates is
\begin{eqnarray}
\frac{\partial}{\partial t} \rho + \nabla\cdot (\rho \vecv) & = & 0 \\
\frac{\partial}{\partial t} \rho + \frac{1}{r^2}\frac{\partial}{\partial r}(r^2 \rho v) & = & 0,
\end{eqnarray}
where $v$ is the radial velocity of the gas.

It is useful to write the equations in terms of $M_r$ rather than $\rho$, so we take the time derivative of $M_r$ to get
\begin{eqnarray*}
\frac{\partial}{\partial t}M_r & = & 4\pi \int_{0}^{r'} r'^2 \frac{\partial}{\partial t} \rho \,dr' \\
& = & -4\pi \int_{0}^{r'} \frac{\partial}{\partial r'}(r'^2 \rho v)\, dr' \\
& = & -4\pi r^2 \rho v \\
& = & -v \frac{\partial}{\partial r}M_r.
\end{eqnarray*}
In the second step we used the mass conservation equation to substitute for $\partial \rho\partial t$, and in the final step we used the definition of $M_r$ to substitute for $\rho$. To figure out how the gas moves, we write down the Navier-Stokes equation without viscosity, which is just the Lagrangean version of the momentum equation:
\begin{equation}
\rho \frac{Dv}{Dt} = -\frac{\partial}{\partial r}P - \mathbf{f}_g,
\end{equation}
where $\mathbf{f}_g$ is the gravitational force. For the momentum equation, we take advantage of the fact that the gas is isothermal to write $P=\rho c_s^2$. The gravitational force is $\mathbf{f}_g = -G M_r / r^2$. Thus we have
\begin{equation}
\frac{Dv}{Dt}= \frac{\partial}{\partial t}v + v\frac{\partial}{\partial r} v= -\frac{c_s^2}{\rho} \frac{\partial}{\partial r}{\rho} - \frac{G M_r}{r^2}.
\end{equation}

For a given set of initial conditions, it is generally very easy to solve these equations numerically, and in some cases to solve them analytically. To get a sense of what to expect, let's think about the behavior in the limit of zero gas pressure, i.e.\ $c_s = 0$. We take the gas to be at rest at $t=0$. This is not as bad an approximation as you might think. Consider the virial theorem: the thermal pressure term is just proportional to the mass, since the gas sound speed stays about constant. On the other hand, the gravitational term varies as $1/R$. Thus, even if pressure starts out competitive with gravity, as the core collapses the dominance of gravity will increase, and before too long the collapse will resemble a pressureless one.

In this case the momentum equation is trivial:
\begin{equation}
\frac{Dv}{Dt} = -\frac{GM_r}{r^2}.
\end{equation}
This just says that a shell's inward acceleration is equal to the gravitational force per unit mass exerted by all the mass interior to it, which is constant. We can then solve for the velocity as a function of position:
\begin{equation}
v = \dot{r} = -\sqrt{2GM_r}\left(\frac{1}{r_0}-\frac{1}{r}\right)^{1/2},
\end{equation}
where $r_0$ is the position at which a particular fluid element starts. To integrate again and solve for $r$, we make the substitution $r=r_0 \cos^2\xi$ \cite{hunter62a}:
\begin{eqnarray}
-2 r_0 (\cos\xi \sin\xi) \dot{\xi} & = & -\sqrt{\frac{2GM_r}{r_0}} \left(\frac{1}{\cos^2\xi}-1\right)^{1/2} \\
2 (\cos\xi\sin\xi) \dot{\xi} & = & \sqrt{\frac{2GM_r}{r_0^3}}\tan\xi \\
2 \cos^2\xi\, d\xi & = & \sqrt{\frac{2GM_r}{r_0^3}} dt \\
\xi+\frac{1}{2}\sin 2\xi & = & t \sqrt{\frac{2GM_r}{r_0^3}}.
\end{eqnarray}

We are interested in the time at which a given fluid element reaches the origin, $r=0$. This corresponds to $\xi = \pi/2$, so this time is
\begin{equation}
t = \frac{\pi}{2}\sqrt{\frac{r_0^3}{2 G M_r}}.
\end{equation}
Suppose that the gas we started with was of uniform density $\rho$, so that $M_r = (4/3)\pi r_0^3 \rho$. In this case we have
\begin{equation}
t = t_{\rm ff} = \sqrt{\frac{3\pi}{32 G \rho}},
\end{equation}
where we have defined the free-fall time $t_{\rm ff}$: it is the time required for a uniform sphere of pressureless gas to collapse to infinite density.

For a uniform fluid this means that the collapse is synchronized -- all the mass reaches the origin at the exact same time. A more realistic case is for the initial state to have some level of central concentration, so that the initial density rises inward. Let's take the initial density profile to be $\rho = \rho_c (r/r_c)^{-\alpha}$, where $\alpha > 0$ so the density rises inward. The corresponding enclosed mass is
\begin{equation}
M_r = \frac{4}{3-\alpha}\pi \rho_c r_c^3 \left(\frac{r}{r_c}\right)^{3-\alpha} 
\end{equation}
Plugging this in, the collapse time is
\begin{equation}
t = \sqrt{\frac{(3-\alpha)\pi}{32 G \rho_c}} \left(\frac{r_0}{r_c}\right)^{\alpha/2}.
\end{equation}
Since $\alpha>0$, this means that the collapse time increases with initial radius $r_0$.

This illustrates one of the most basic features of a collapse, which will continue to hold even in the case where the pressure is non-zero. Collapse of centrally concentrated objects occurs inside-out, meaning that the inner parts collapse before the outer parts. Within the collapsing region near the star, the density profile also approaches a characteristic shape. If the radius of a given fluid element $r$ is much smaller than its initial radius $r_0$, then its velocity is roughly
\begin{equation}
v \approx v_{\rm ff}\equiv -\sqrt{\frac{2GM_r}{r}},
\end{equation}
where we have defined the free-fall velocity $v_{\rm ff}$ as the characteristic speed achieved by an object collapsing freely onto a mass $M_r$.

The mass conservation equation is
\begin{equation}
\frac{\partial M_r}{\partial t} = -v\frac{\partial M_r}{\partial r}  = -4\pi r^2 v \rho
\end{equation}
If we are near the star so that $v\approx v_{\rm ff}$, then this implies that
\begin{equation}
\rho = \frac{(\partial M_r/\partial t) r^{-3/2}}{4\pi\sqrt{2 G M_r}}.
\end{equation}
To the extent that we look at a short interval of time, over which the accretion rate does not change much (so that $\partial M_r /\partial t$ is roughly constant), this implies that the density near the star varies as $\rho\propto r^{-3/2}$.

\paragraph{The characteristic accretion rate}

What sort of accretion rate do we expect from a collapse like this? For a core of mass $M_c = [4/(3-\alpha)]\pi \rho_c r_c^3$, the last mass element arrives at the center at a time
\begin{equation}
t_c = \sqrt{\frac{(3-\alpha)\pi}{32 G \rho_c}} = \sqrt{\frac{3-\alpha}{3}}t_{\rm ff}(\rho_c),
\end{equation}
so the time-averaged accretion rate is
\begin{equation}
\langle\dot{M}\rangle = \sqrt{\frac{3}{3-\alpha}} \frac{M_c}{t_{\rm ff}(\rho_c)}.
\end{equation}

In order to get a sense of the numerical value of this, let us suppose that our collapsing object is a marginally unstable Bonnor-Ebert sphere, with mass
\begin{equation}
M_{\rm BE} = 1.18 \frac{c_s^4}{\sqrt{G^3 P_s}},
\end{equation}
where $P_s$ is the pressure at the surface of the sphere and $c_s$ is the thermal sound speed in the core. Let's suppose that the surface of the core, at radius $r_c$, is in thermal pressure balance with its surroundings. Thus $P_s = \rho_c c_s^2$, so we may rewrite the Bonnor-Ebert mass as
\begin{equation}
M_{\rm BE} = 1.18 \frac{c_s^3}{\sqrt{G^3 \rho_c}}.
\end{equation}

A Bonnor-Ebert sphere doesn't have a powerlaw structure, but if we substitute into our equation for the accretion rate and say that the factor of $\sqrt{3/(3-\alpha)}$ is a number of order unity, we find that the accretion rate is
\begin{equation}
\langle\dot{M}\rangle \approx \frac{c_s^3/\sqrt{G^3\rho_c}}{1/\sqrt{G\rho_c}} = \frac{c_s^3}{G}.
\end{equation}
This is an extremely useful expression, because we know the sound speed $c_s$ from microphysics. Thus, we have calculated the rough accretion rate we expect to be associated with the collapse of any object that is marginally stable based on thermal pressure support. Plugging in $c_s=0.2$ km s$^{-1}$, we get $\dot{M} \approx 2\times 10^{-6}$ $\msun$ yr$^{-1}$ as the characteristic accretion rate for these objects.

Since the typical stellar mass is a few tenths of $\msun$, based on the peak of the IMF, this means that the characteristic star formation time is of order $10^5-10^6$ yr. Of course this conclusion about the accretion rate only applies to collapsing objects that are supported mostly by thermal pressure. Other sources of support produce higher accretion rates; this is typically the case for massive stars.

\subsubsection{Rotation Collapse and the Angular Momentum Problem}

The next element to add to this picture is rotation. We characterize the importance of rotation through the ratio of rotational kinetic energy to gravitational binding energy, which we denote $\beta$. If the angular velocity of the rotation is $\Omega$ and the moment of inertia of the core is $I$, this is
\begin{equation}
\beta = \frac{(1/2)I\Omega^2}{a G M^2/R},
\end{equation}
where $a$ is our usual numerical factor that depends on the mass distribution. For a sphere of uniform density $\rho$, we get
\begin{equation}
\beta = \frac{1}{4\pi G \rho}\Omega^2 = \frac{\Omega^2 R^3}{3 G M}
\end{equation}
Thus we can estimate $\beta$ simply given the density of a core and its measured velocity gradient. Observed values of $\beta$ are typically a few percent \cite{goodman93a}.

Let us consider how rotation affects the collapse, for a simple core of constant angular velocity $\Omega$. Consider a fluid element that is initially at some distance $r_0$ from the axis of rotation. We will consider it to be in the equatorial plane, since fluid elements at equal radius above the plane have less angular momentum, and thus will fall into smaller radii. Its initial angular momentum in the direction along the rotation axis is $j=r_0^2\Omega$.

If pressure forces are insignificant for this fluid element, it will travel ballistically, and its specific angular momentum and energy will remain constant as it travels. At its closest approach to the central star plus disk, its radius is $r_{\rm min}$ and by conservation of energy its velocity is $v_{\rm max} = \sqrt{2 G M_*/r_{\rm min}}$, where $M_*$ is the mass of the star plus the disk material interior to this fluid element's position. Conservation of angular momentum them implies that $j=r_{\rm min} v_{\rm max}$.

Combining these two equations for the two unknowns $r_{\rm min}$ and $v_{\rm max}$, we have
\begin{equation}
r_{\rm min} = \frac{r_0^4 \Omega^2}{G M_*} = \frac{4\pi \rho \beta r_0^4}{M_*},
\end{equation}
where we have substituted in for $\Omega^2$ in terms of $\beta$. This tells us the radius at which infalling material must go into a disk because conservation of angular momentum and energy will not let it get any closer.

We can equate the stellar mass $M_*$ with the mass that started off interior to this fluid element's position -- this amounts to assuming that the collapse is perfectly inside-out, and that the mass that collapses before this fluid element's all makes it onto the star. If we make this approximation, then $M_*=(4/3)\pi \rho r_0^3$, and we get
\begin{equation}
r_{\rm min} = 3 \beta r_0,
\end{equation}
i.e.\ the radius at which the fluid element settles into a disk is simply proportional to $\beta$ times a numerical factor of order unity.

We shouldn't take the factor too seriously, since of course real clouds aren't uniform spheres in solid body rotation, but the result that rotation starts to influence collapse and force disk formation at a radius that is a few percent of the core radius is interesting. It implies that for cores that are $\sim 0.1$ pc in size and have $\beta$ values typical of what is observed, they should start to become rotationally flattened at radii of several hundred AU. This will be the typical size scale of protostellar disks.

In order for mass to actually get to a star, of course, its angular momentum must be redistributed outwards. It must get from hundreds of AU to $\ll 1$ AU. Fortunately, disks are devices whose sole purpose is to separate mass and angular momentum. We will not spend any more time on disks (which could form an entire lecture series of their own), except to say that they provide numerous possible mechanisms to remove the angular momentum from the bulk of the mass and allow it to reach the star.

\subsubsection{Magnetized Collapse and the Magnetic Flux Problem}

So far we have only dealt with pressure, rotation, and gravity. Now we will add magnetic fields to the picture. We will assume that we have a magnetically supercritical core, so that we need not worry about magnetic fields significantly inhibiting the collapse. Instead, we will work on a second problem: that of the magnetic flux.

As we discussed earlier, observed magnetic fields make cores marginally supercritical, but only by factors of a few. If the collapse occurs in the ideal-MHD regime, where perfect flux-freezing holds, then this mass to flux ratio doesn't change. What sort of magnetic field would we then expect stars to have? For the Sun, if we had $M_\Phi = M/2$, then we would expect the mean magnetic field to be
\begin{equation}
\Phi_B = \pi \rsun^2 B = \frac{G^{1/2} \msun}{0.24}
\quad \Longrightarrow \quad
B = \frac{G^{1/2} \msun}{0.24\pi \rsun^2} \approx 10^8\mbox{ G}
\end{equation}

For comparison, the observed mean surface magnetic field of the Sun is a few Gauss. Clearly this means that the Sun, and other stars like it, must have lost most of their magnetic fields during the collapse process. This means that the ideal MHD regime cannot apply, and resistivity or some other non-ideal effect must become significant.

There are two mechanisms which can lead to violation of flux-freezing in cores: ambipolar diffusion and Ohmic resistivity. As we saw in the last section, ambipolar diffusion will cause ions and neutrals to begin decoupling on scales below $\sim 0.5$ pc.

Decoupling does not prevent the field from increasing at all -- there is always some inward drag exerted on the ions by the infalling neutrals, even if it is weak. This will eventually increase the field strength, which leads to the second effect: Ohmic resistivity. As the field lines are pressed closer together, field lines of opposite direction come into close proximity. When this happens, the field can reconnect, meaning that its topology changes and drops to a lower energy state. The excess energy is released in the form of heat. The microphysics of this process is not fully understood, but we see it happening in plasmas like the solar corona, where it is associated with flaring. Something similar must happen in protostellar cores in order to explain the observed low magnetic fields of stars.

\section{Two Problems: The Star Formation Rate and the Initial Mass Function}

In this final section we'll come up to the present state of the art and talk about what are probably the two largest unsolved problems in star formation today: the star formation rate and the origin of the initial mass function.

\subsection{The Star Formation Rate}

\subsubsection{The Observational Problem: Slow Star Formation}

The problem of the star formation rate can be understood very simply. In the last lecture we computed the characteristic timescale for collapse to occur, and argued that, even if a collapsing region is only slightly unstable initially, this will not change the collapse time by much. Magnetic fields could delay or prevent collapse, but observations seem to indicate that they are not strong enough to do so. Thus we would expect that, on average, clouds will collapse on a timescale comparable to $t_{\rm ff}$, and the rate of star formation in a galaxy should be the total mass of bound molecular clouds $M$ divided by this.

To make this more concrete, we introduce the notation (first used by Krumholz \& McKee \cite{krumholz05c})
\begin{equation}
\eff = \frac{\dot{M}_*}{M(\rho)/t_{\rm ff}(\rho)},
\end{equation}
where $M(\rho)$ is the gas mass in a given region with density $\rho$ or larger, $t_{\rm ff}$ is the free-fall time evaluated at that density, and $\dot{M}_*$ is the star formation rate in the region in question. The regions here can be either entire galaxies are specified volumes within a galaxy. We refer to $\eff$ as the dimensionless star formation rate or star formation efficiency. Unfortunately the language here is somewhat confused, because people sometimes mean something different by star formation efficiency. To avoid confusion we will just use the symbol $\eff$.

The argument we have just given suggests that $\eff$ should be of order unity if we pick $\rho$ to be the typical density of molecular clouds, or anything higher. However, the actual value of $\eff$ is much smaller, as first pointed out by Zuckerman \& Evans \cite{zuckerman74a}. The Milky Way's disk contains $\sim 10^9$ $\msun$ of GMCs inside the Solar circle \cite{williams97a, bronfman00a}, and these have a mean density of $n\sim 100$ cm$^{-3}$ \cite{solomon87a}, corresponding to a free-fall time time $t_{\rm ff} \approx 4$ Myr. Thus $M/t_{\rm ff} \approx 250$ $\msun$ yr$^{-1}$. The observed star formation rate in the Milky Way is $\sim 1$ $\msun$ yr$^{-1}$ \cite{mckee97a, murray10b}. Thus $\eff \sim 0.01$! Clearly our naive estimate is wrong.

\begin{figure}
\includegraphics[height=.3\textheight]{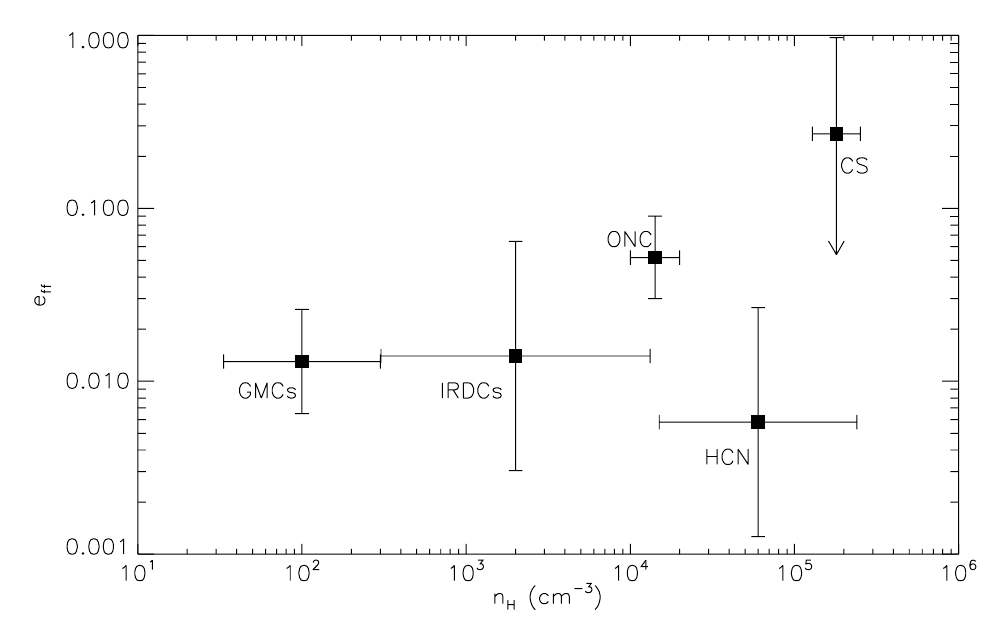}
  \caption{Observed star formation efficiency per free-fall time $\eff$, as a function of mean gas density $n_{\rm H}$. Each data point represents a different method of measuring the gas, which is sensitive to different densities. GMC indicates giant molecular clouds, traced in CO $J=1\rightarrow 0$. IRDCs indicates infrared dark clouds, measured in infrared absorption. ONC is the Orion Nebula cluster, a single star cluster near Earth, whose gas mass is estimated from the mass and dynamical state of the remaining star cluster. HCN represents extragalactic measurements in the HCN $J=1\rightarrow 0$ line. Finally, CS represents measurements of the CS $J=5\rightarrow 4$ line within the Milky Way. However, it is only an upper limit. Reprinted with permission from the AAS from \cite{krumholz07e}.
  \label{fig:eff}}
\end{figure}

One can repeat this exercise in many galaxies and using many different density tracers. One way is to measure the mass using a molecular tracer with a known critical density, which effectively gives $M(\rho)$, compute $t_{\rm ff}(\rho)$ at that critical density, and compare to the star formation rate. Krumholz \& Tan \cite{krumholz07e} compiled the data available at the time and found that, for every tracer for which they could make a measurement, and in every galaxy, $\eff$ was still $\sim 0.01$ (Figure \ref{fig:eff}. Subsequent more accurate measurements in several large surveys, most notably the c2d survey \cite{evans09a}, give the same result. Thus, we have a problem: why is the star formation rate about 1\% of the naively estimated value?

As a side note, the famous Kennicutt relation \cite{kennicutt98a}, which is an observed correlation between the star formation rate in a given portion of a galaxy disk and the gas surface density in that region. The observed normalization of the Kennicutt relation is equivalent to the statement $\eff\sim 0.01$.

\subsubsection{Potential Solutions}

There are two major classes of proposed solution to this problem. One is the idea that molecular clouds aren't really gravitationally bound, and the other is the idea that clouds are bound, but that turbulence inhibits large-scale collapse while permitting small amount of mass to collapse.

\paragraph{Unbound GMCs}

The unbound GMC idea is that most of the mass in molecular clouds is in a diffuse state that is either not gravitationally bound, or that is supported against collapse by strong magnetic fields. (The latter is not ruled out because it is not easy to measure the magnetic field in very diffuse molecular gas.) This idea would definitely work, in the sense that it would produce low star formation rates, if GMCs really were unbound. The main problem is that there is no observational evidence that this is the case, and considerable evidence that it is not. In particular, while we have no trouble finding low mass CO-emitting clouds with virial ratios $\avir \gg 1$, CO-emitting clouds with masses $\gtsim 10^4$ $\msun$ and $\avir \gg 1$ do not appear to exist \cite{heyer01a}. If GMCs were really unbound, why do they all have virial ratios $\sim 1$?

A second problem with this idea is that, as we have seen $\eff$ is $\sim 1\%$ across of huge range of densities and environments. It is not at all obvious why the fraction of mass that is bound would be the same at all densities and across all galactic environments, from low-mass dwarfs to massive ultraluminous infrared galaxies. The universality of the $\sim 1\%$ seems to demand an explanation that is rooted in something more universal than an appeal to fractions of a GMC that are bound versus unbound.

\paragraph{Turbulence-Regulated Star Formation}

A more promising idea, which is probably the most generally accepted at this point (though it still has significant problems) is that the ubiquitous turbulence observed in GMCs serves to keep the star formation rate within them low. The first quantitative model of this in the hydrodynamic case was proposed by Krumholz \& McKee \cite{krumholz05c}, and it has since been extended to the MHD case by Padoan \& Nordlund \cite{padoan09a}.

The basic idea of this model relies on two properties of supersonic turbulence. Due to time limitations we will not prove these, but they can be understood analytically, and the are reproduced in every simulation. The first property is that turbulence obeys what is known as a linewidth-size relation. This means that, if we consider a region of size $\ell$ and compute the non-thermal velocity dispersion $\sigma_{\rm nt}$ within it, the velocity dispersion will depend on $\ell$. For subsonic turbulence the relationship is $\sigma_{\rm nt} \propto \ell^{1/3}$, while for highly supersonic turbulence it is $\sigma\propto \ell^{1/2}$. This relationship is in fact observed in molecular clouds. 

Now consider the implications of this result in the virial theorem. On large scales we know that clouds have $\avir\sim 1$, so that $|\mathcal{W}| \sim \mathcal{T}$. If we consider a random region within a cloud of size $\ell < R$, where $R$ is the cloud radius, then the mass within that region will scale as $\ell^3$, so the gravitational potential energy will scale as $\mathcal{W} \propto M^2/\ell \propto \ell^5$. In comparison, the kinetic energy varies as $\mathcal{T} \propto M\sigma^2 \propto \ell^4$, since $\sigma^2\propto \ell$ for large $\ell$. Thus we expect that, for an average region
\begin{equation}
\avir(\ell) \propto \frac{\mathcal{T}}{\mathcal{W}} \propto \frac{1}{\ell}.
\end{equation}
Since $\avir(R) \approx 1$, this means that $\avir \gg 1$ for $\ell \ll R$, i.e.\ the typical, randomly chosen region within a GMC is gravitationally unbound by a large margin. This is in good agreement with observations: GMCs are bound, but random sub-regions within them are not.

We can turn this around by asking how much denser than average a region must be in order to be bound. For convenience we define the sonic length as the choice of length scale $\ell$ for which the non-thermal velocity dispersion is equal to the thermal sound speed, i.e.\
\begin{equation}
\sigma = c_s \left(\frac{\ell}{\lambda_s}\right)^{1/2}.
\end{equation}
Now consider a region within a cloud with density $\rho$, chosen small enough that the velocity dispersion is dominated by thermal rather than non-thermal motions. The maximum mass that can be supported against collapse by thermal pressure is the Bonnor-Ebert mass. If we let $\rho$ be the density at the surface of our Bonnor-Ebert sphere and we adopt a uniform sound speed $c_s$, then $\sigma=c_s$, $P_S = \rho c_s^2$, and
\begin{equation}
M_{\rm BE} =  1.18 \frac{c_s^3}{\sqrt{G^3\rho}},
\end{equation}
and the corresponding radius of the maximum mass sphere is
\begin{equation}
R_{\rm BE} = 0.66 \frac{c_s}{\sqrt{G\rho}}.
\end{equation}

We can compute the gravitational potential energy and the thermal energy of such a sphere from its self-consistently determined density distribution. The result is
\begin{eqnarray}
\mathcal{W} & = & -1.06 \frac{c_s^5}{\sqrt{G^3 \rho}} \\
\mathcal{T}_{\rm th} & = & 1.14 |\mathcal{W}|.
\end{eqnarray}
Similarly, we can compute the turbulent energy from the linewidth-size relation evaluated at $\ell=2 R_{\rm BE}$. Doing so we have
\begin{equation}
\mathcal{T}_{\rm turb} = \frac{3}{2} M_{\rm BE} \sigma^2(2R_{\rm BE}) = 0.89 \left(\frac{\lambda_J}{\lambda_s}\right) |\mathcal{W}|,
\end{equation}
where $\lambda_J = \sqrt{\pi c_s^2/G\rho}$ is called the Jeans length; it is just $2.7$ times $R_{\rm BE}$.

This is a very interesting result. It says that the turbulent energy in a maximal-mass Bonnor Ebert sphere is comparable to its gravitational potential energy if the Jeans length is comparable to the sonic length. Since the Jeans length goes up as the density goes down, this means that, at low density, $\mathcal{T}_{\rm turb} \gg |\mathcal{W}|$, while at high density $\mathcal{T}_{\rm turb} \gg |\mathcal{W}|$. In order for a region to be unstable to collapse, the latter condition must hold. We have therefore identified a minimum density at which we expect sub-regions of a molecular cloud to be unstable to collapse.

To get a sense of what this density is, let us evaluate $\lambda_J/\lambda_s$ at the mean density of a $10^4$ $\msun$, 6 pc-sized molecular cloud. If such a cloud has $\avir=1$, the velocity dispersion at the cloud scale is $\sigma=1.2$ km s$^{-1}$; since $c_s=0.2$ km s$^{-1}$, we have $\lambda_s=0.15$ pc. At the mean density of the cloud, $\rho=7.5\times 10^{-22}$ g cm$^{-3}$, and $\lambda_J = 1.5$ pc. Thus $\lambda_J \gg \lambda_s$, and at the mean density things are unbound by a large margin. To be dense enough to be bound, the density has to be larger than the mean by a factor of $(\lambda_s/\lambda_J)^2\approx 100$, so bound structures are those with $\rho \gtsim 8\times 10^{-20}$ g cm$^{-3}$, or $n \gtsim 3\times 10^4$ cm$^{-3}$.

In order to go further we must know something about the internal density distribution in molecular clouds. We now invoke the second property of supersonic isothermal turbulence: it generates a distribution of densities that is lognormal in form. Formally, the point probability distribution function of the density, meaning the probability of measuring a density $\rho$ at a given position, obeys
\begin{equation}
\frac{dp}{dx} = \frac{1}{2\pi \sigma_\rho^2} \exp\left[-\frac{\left(\ln x - \overline{\ln x}\right)^2}{2\sigma_\rho^2}\right],
\end{equation}
where $x=\rho/\overline{\rho}$ is the density divided by the volume-averaged density, $\overline{\ln x} = -\sigma_\rho^2/2$ is the mean of the logarithm of the overdensity, and $\sigma_\rho$ is the dispersion of log density. That the density distribution should be lognormal isn't surprising. In a supersonically turbulent medium, each shock that passes a point multiplies its density by a factor of the Mach number of the shock squared, and each rarefaction front divides the density by a similar factor. Thus the density at a point is a product of many multiplications and divisions, and by the central limit theorem the result of many such operations is a lognormal (just as the result of doing many random additions and subtractions is a normal distribution). Empirical work shows that the width of the normal distribution depends on the Mach number $\mathcal{M}$ of the turbulence as
\begin{equation}
\sigma_\rho \approx \left[\ln\left(1 + \frac{3\mathcal{M}^2}{4}\right)\right]^{1/2}.
\end{equation}

Now we can put together an estimate of the star formation rate. We estimate that the gas that has a density larger than the critical density given by the condition that $\lambda_J < \lambda_s$, which is
\begin{equation}
x_{\rm crit} = \left(\phi_x \frac{\lambda_J(\overline{\rho})}{\lambda_s}\right)^2,
\end{equation}
where $\phi_x$ is a factor of order unity. If we compute the mean density and the sonic length for our fiducial cloud of mass $M$, radius $R$, and velocity dispersion $\sigma$, with a little algebra we can show that
\begin{equation}
x_{\rm crit} = \frac{\pi^2 \phi_x^2}{15} \avir \mathcal{M}^2.
\end{equation}
Gas above this density forms stars on a timescale given by the free-fall time at the mean density, since that is the timescale over which the density distribution will be regenerated to replace overdense regions that collapse to stars. Thus we have
\begin{eqnarray}
\epsilon_{\rm ff} & = & \phi \int_{x_{\rm crit}}^\infty x \frac{dp}{dx} \, dx \\
& = & \frac{\phi}{2} \left[1 + \mbox{erf}\left(\frac{\sigma_\rho^2-2\ln x_{\rm crit}}{2^{3/2} \sigma_\rho}\right)\right],
\end{eqnarray}
where $\phi$ is another constant of order unity. Note that, except for the $\phi$ factors, everything in this expression is given in terms of $\avir$ and $\mathcal{M}$, i.e.\ in terms of the virial ratio and Mach number of the cloud. The $\phi$ factors can be calibrated against simulations. For those who don't walk around with graphs of the the error function in their heads (i.e.\ most of us), it's useful to have a powerlaw approximation to this, which is
\begin{equation}
\eff \approx 0.017 \avir^{-0.68} (\mathcal{M}/100)^{-0.32}.
\end{equation}
In other words, for a cloud with $\mathcal{M} \sim 10-100$ and $\avir = 1$, we expect $\eff \sim 0.01$, which nicely explains the observation that $\eff\sim 0.01$ everywhere. This analytic model also agrees well with numerical simulations \cite{padoan09a}.

\paragraph{Driving Turbulence}

This is a cute explanation, but it assumes that the turbulence is present and is capable of inhibiting star formation over the lifetime of a molecular cloud. This is not obvious, because we know from numerical experiments that turbulence decays quickly. This is not surprising. Every time there is a shock, kinetic energy is converted into thermal energy. Because radiative times are short compared to mechanical ones, as we showed earlier, all this energy is radiated away immediately, bringing the gas back to its original temperature. This represents a net loss of energy, and in the absence of a source to offset this loss the turbulence must decay. Numerical experiments show that the decay time is only about a crossing time of the cloud \cite{stone98a}.

We therefore need an energy source to drive the turbulence. There are two main possibilities, both of which probably contribute. One is the gravitational potential energy released in the formation of the molecular cloud itself. As material falls onto the cloud it can drive turbulent motion, and as long as the cloud gains mass quickly from the larger ISM that is probably an important energy source \cite{vazquez-semadeni10a, goldbaum10a}. A second source, which is probably more important in evolved GMCs, is feedback from newly formed stars. Young stars produce strong jets that can drive motions in their parent clouds \cite{li06b, wang10a}, and they also produce ionizing radiation that can drive motions \cite{krumholz06d, krumholz09d}. Both of these effects can drive turbulence, and they probably dominates in more evolved clouds. Exactly what the energy balance in GMCs is, and how it is maintained, is not completely understood.

\subsection{The Initial Mass Function}

\subsubsection{The Observed IMF}

Our second unsolved problem in this lecture is the initial mass function (IMF). To begin, we have to define what the IMF means. It is simply the mass distribution of a population of stars at birth. We define this by a function
\begin{equation}
\xi(m) = \frac{dn}{d\ln m}.
\end{equation}
Note that $dn/dm$ would be the number of stars per unit mass, while $\xi(m) = dn/d\ln m = m (dn/dm)$ is the mass of stars per unit mass, i.e.\ $\int_{m_1}^{m_2} \xi(m) \, dm$ is the fraction of the mass in a newborn stellar population that is found in stars with masses between $m_1$ and $m_2$.

Observing the IMF is tricky, and there are three main approaches. One is to look at a young cluster and count the stars in it as a function of mass. This is the most straightforward approach, but it is limited by the number of young clusters where we can directly measure individual stars down to low masses. This means that we get a clean measurement, but the statistics are poor. A second approach is to rely on counts of field stars in the solar neighborhood that are no longer in clusters. Here the statistics are much better, but we can only use this technique for low mass stars, because for massive ones the number in the Solar neighborhood is determined more by star formation history than by the IMF. Finally, we can get limits on the IMF from the integrated light of a stellar populations.

\begin{figure}
\includegraphics[height=.3\textheight]{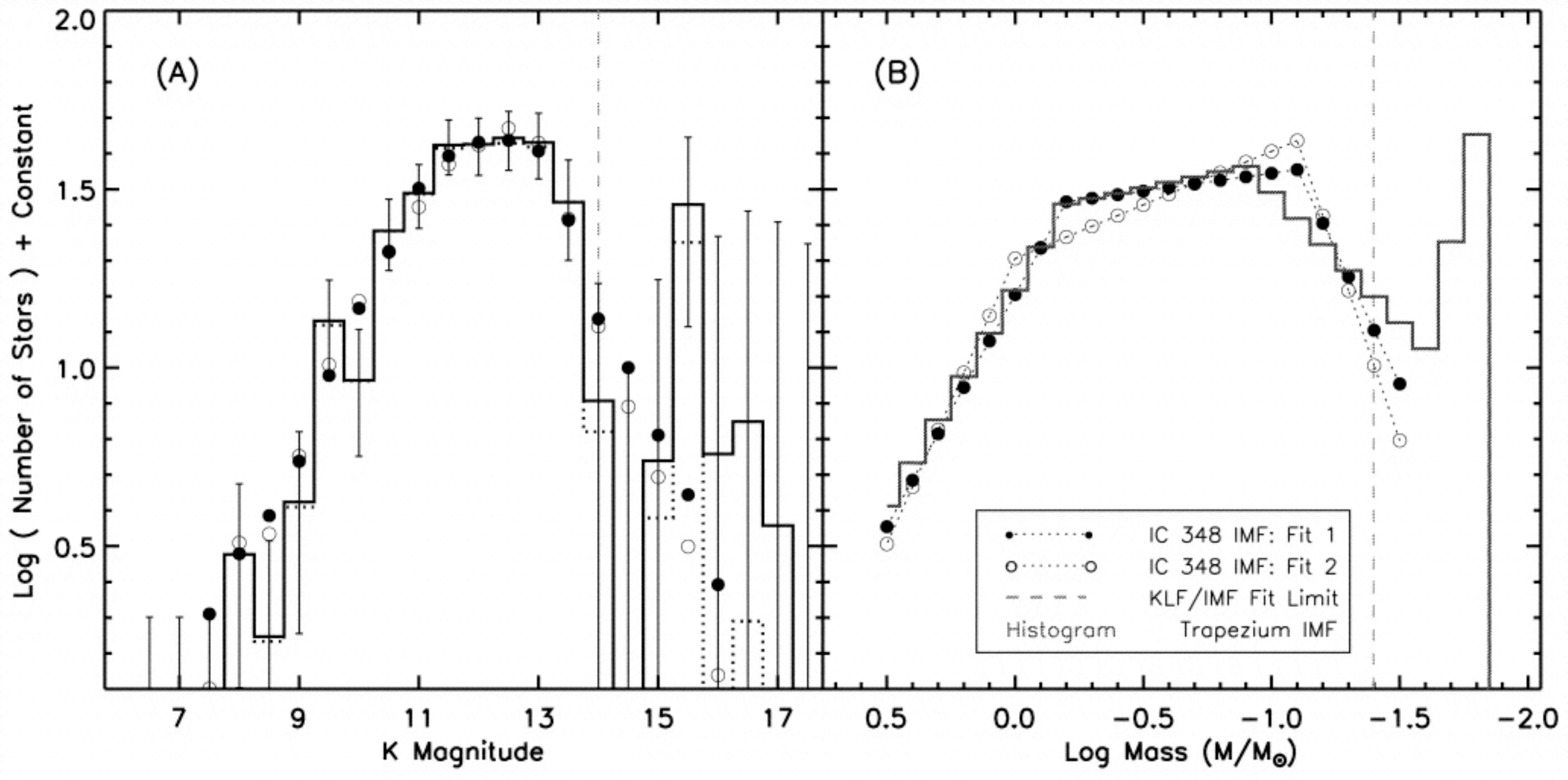}
  \caption{Measured K band luminosity functions (left) and stellar initial mass functions (right) for the cluster IC 348 (filled and open circles), and for the Trapezium cluster (histogram). Reprinted with permission from the AAS from \cite{muench03a}.
  \label{fig:imf}}
\end{figure}

One interesting result to come out of all of this work is that the IMF is remarkably uniform.  One can notice this at first by comparing the mass distributions of stars in different clusters. As Figure \ref{fig:imf} illustrates, clearly the two clusters IC 348 and the Trapezium have the same IMF for the mass range they cover. This is despite the fact that the Trapezium is a much larger, denser cluster forming out a significantly larger molecular cloud. The Trapezium IMF is also a good fit in a remarkably broad range of even more different environments. For example, it is a good fit to the stellar mass distribution in the Digel 2 North and South clusters, which are forming in the extreme outer galaxy, $R_{\rm gal}\approx 19$ kpc \cite{yasui08a}. We also obtain a good fit using this IMF to model globular clusters, provided that we account for the age of the stellar population and for dynamical effects such as evaporation and mass segregation \cite{de-marchi00a}. This represents a star-forming environment that is much denser, at much lower metallicity, out of the galactic plane rather than in the plane, and at much higher redshift, yet has the same IMF. All of these IMFs also agree with the IMF derived for field stars in the solar neighborhood. Thus one constraint on theories of the IMF is that, at least on the scale of star clusters or larger, it is remarkably universal.
There is some indirect evidence for variation of the IMF at the very high end, although I would describe it as suggestive rather than definitive, and we won't go into it.

\begin{figure}
\includegraphics[height=.3\textheight]{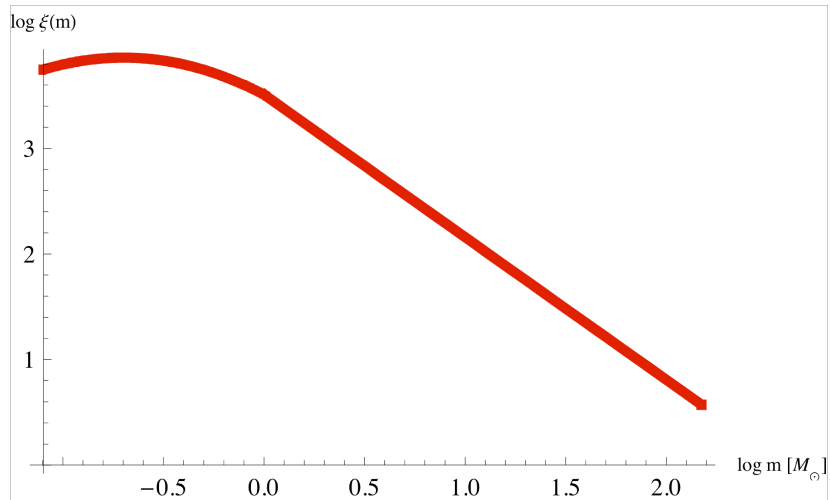}
  \caption{An IMF $\xi(m)$ following the functional form proposed by \cite{chabrier03a}.
  \label{fig:imf2}}
\end{figure}

The observed IMF can be parameterized in several ways; popular parameterizations are due to Kroupa (2002) and Chabrier et al. (2003). All parameterizations share in common that they have a powerlaw tail at high masses with
\begin{equation}
\xi(m) \propto m^{-\Gamma},
\end{equation}
with $\Gamma \approx 1.3-1.4$. At lower masses there is a flattening, reaching a peak around $\sim 0.2-0.3$ $\msun$, and then a decline at still lower masses, although that is very poorly determined due to the difficulty of finding low mass stars. This is parameterized either with a series of broken powerlaws or with a lognormal function \cite{chabrier03a, chabrier05a, kroupa02b}. Figure \ref{fig:imf2} shows a plot of one proposed functional form for $\xi(m)$ (due to \cite{chabrier03a}).

\subsubsection{The IMF in the Gas Phase?}

What is the origin of this universal mass function? The biggest breakthrough in answering this question in the last several years has come not from theoretical or numerical advances (although those have certainly helped), but from observations. In particular, the advent of large-scale mm and sub-mm surveys of star-forming regions has made it possible to assemble statistically significant samples of overdense regions, or ``cores", in star-forming clouds.

\begin{figure}
\includegraphics[height=.3\textheight]{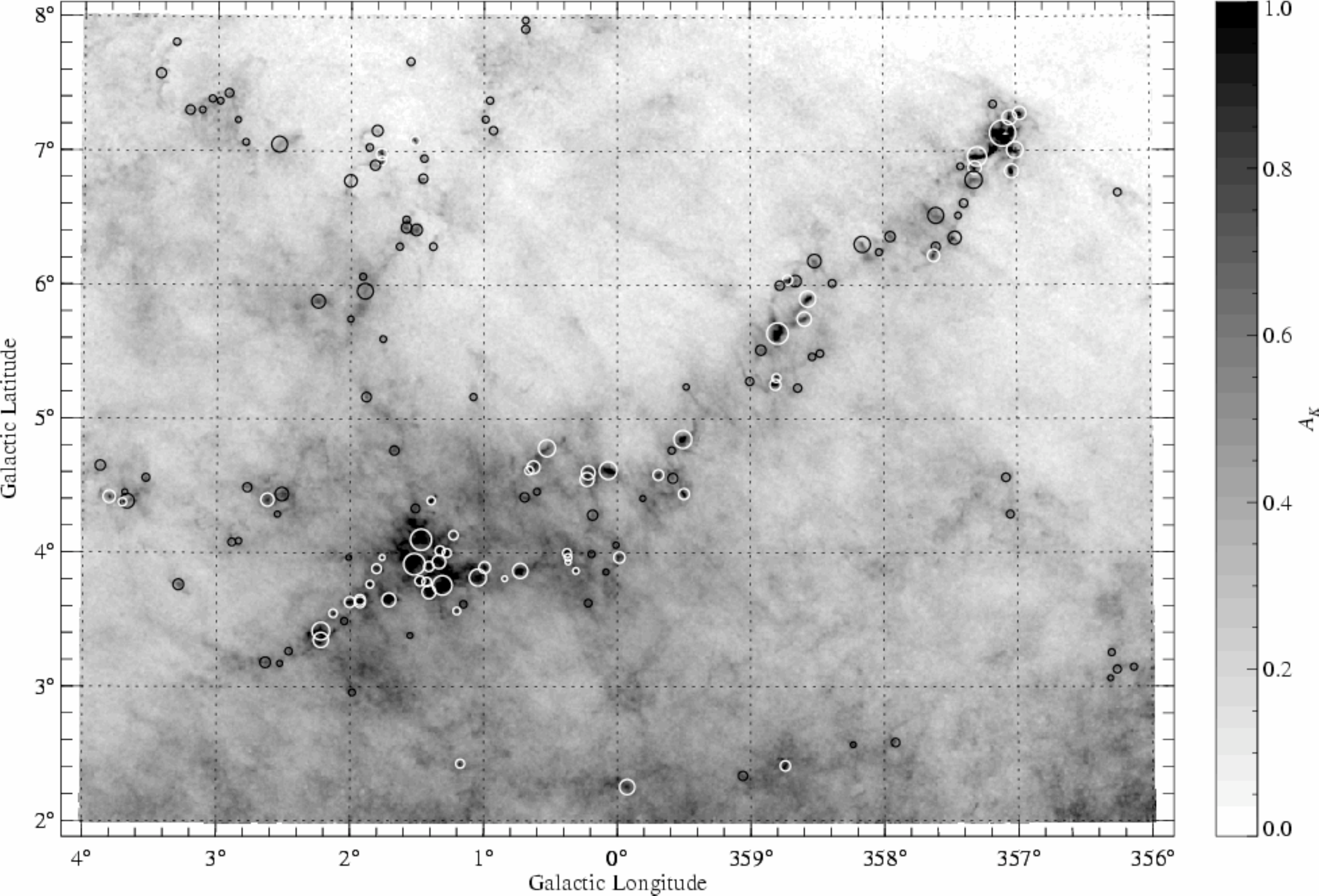}\includegraphics[height=.3\textheight]{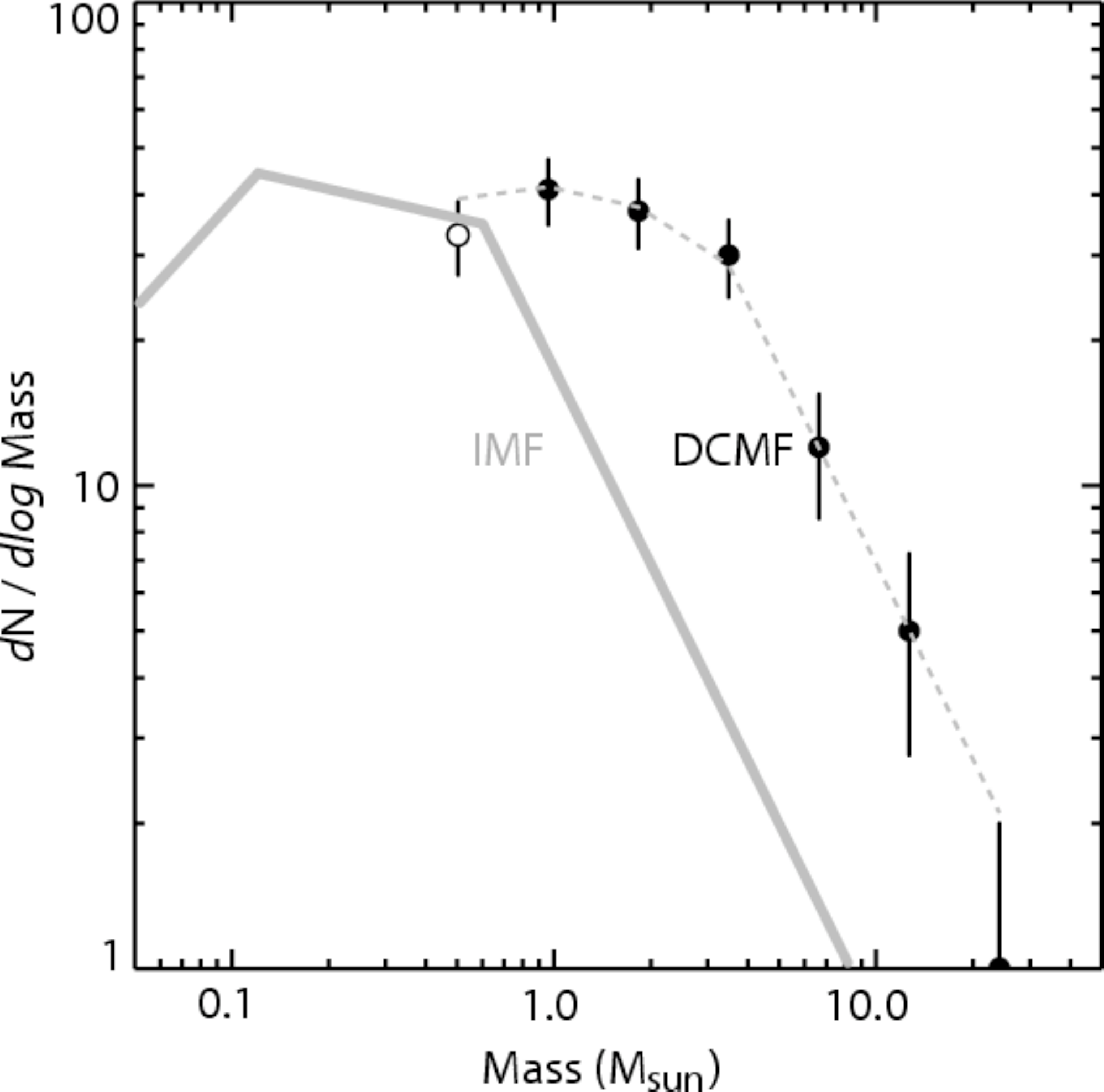}
  \caption{Left: extinction map of the Pipe Nebula with the cores circled. Right: mass function of the cores (data points with error bars) compared to stellar IMF (solid line). Reprinted with permission from \cite{alves07a}.
  \label{fig:pipecores}}
\end{figure}

The remarkable result of these surveys, repeated using many different techniques in many different regions, is that the core mass function (CMF) is the same as the IMF, just shifted to slightly higher masses. The cleanest example of this comes from the Pipe Nebula, where Alves et al.~\cite{alves07a} used a near-infrared extinction mapping technique to locate all the cores down to very low mass limits at very high spatial resolution. Finding all the cores in the Pipe shows that their mass function matches the IMF, including a powerlaw slope of $-1.35$ at the high end, a flattening at lower masses, and a turn-down below that. The distribution is shifted to higher mass than the IMF by a factor of 3 (Figure \ref{fig:pipecores}). This is the cleanest example, but it is not the only one. Similar results are obtained using dust emission in the Perseus, Serpens, and Ophiuchus clouds \cite{enoch08a}.

The strong inference from these observations is that, whatever mechanism is responsible for setting the stellar IMF, it acts in the gas phase, before the stars form. In other words, the IMF is simply a translation of the CMF, with only $\sim 1/3$ of the material in a given core making it onto a star, and the rest being ejected. That ejection fraction is a plausible result of protostellar outflows, as we'll discuss next week. This doesn't by itself represent a theory of the IMF, since it immediately leads to the question of what physical process is responsible for setting the CMF. It does, however, provide an important constraint that what we should be trying to do is to solve three problems: (1) what is responsible for setting the CMF, (2) why is it that cores generally form single stars or star systems regardless of mass, and (3) what sets the efficiency of $\sim 1/3$.

\subsubsection{A Possible Model: Turbulent Fragmentation and Radiation-Suppressed Fragmentation}

Although we don't have a complete model that meets the three conditions outlined above, we can sketch out the beginnings of one. This may be entirely wrong, but it's the idea that, right now, I consider the most promising.

\paragraph{The Padoan \& Nordlund Model for the CMF}

The model's basic elements were originally proposed by Padoan \& Nordlund \cite{padoan02a}. The first element is the idea that supersonic turbulence generates a spectrum of structures with a slope that looks similar to the high end slope of the stellar IMF. Formally, one can show (though we will not in this lecture) that the distribution of fragment masses follows a distribution
\begin{equation}
\frac{dn_{\rm frag}}{d\ln m} \propto m^{3/(4-\beta)},
\end{equation}
where $\beta$ is a numerical factor related to the exponent $q$ in the linewidth-size relation by $\beta = 2q+1$. (Formally $\beta$ is the index of the turbulent power spectrum, so if we have a linewidth-size relation $\Delta v\propto \ell^q$, then one can show that the power spectrum is $P(k)\propto k^\beta$.) To remind you, for subsonic turbulence $q\approx 1/3$ and for highly supersonic turbulence $q\approx 1/2$, corresponding to $\beta=5/3$ or $\beta=2$, respectively. At the Mach numbers in molecular clouds $\beta$ tends to be a bit less than $2$, around $1.9$, giving a slope around $1.4$. Notice that this is very similar to the observed high-mass slope $\Gamma \sim 1.3-1.4$.

By itself this is just a pure powerlaw. However, not all of the structures generated by the turbulence are gravitationally bound and liable to collapse. The very massive ones almost certainly are, because their masses are larger than the Bonnor-Ebert mass for any plausible surface pressure. However, the low mass ones are bound only if they find themselves in regions of high pressure, which lowers the Bonnor-Ebert mass to a value smaller than the mean in the cloud.

\begin{figure}
\includegraphics[height=.3\textheight]{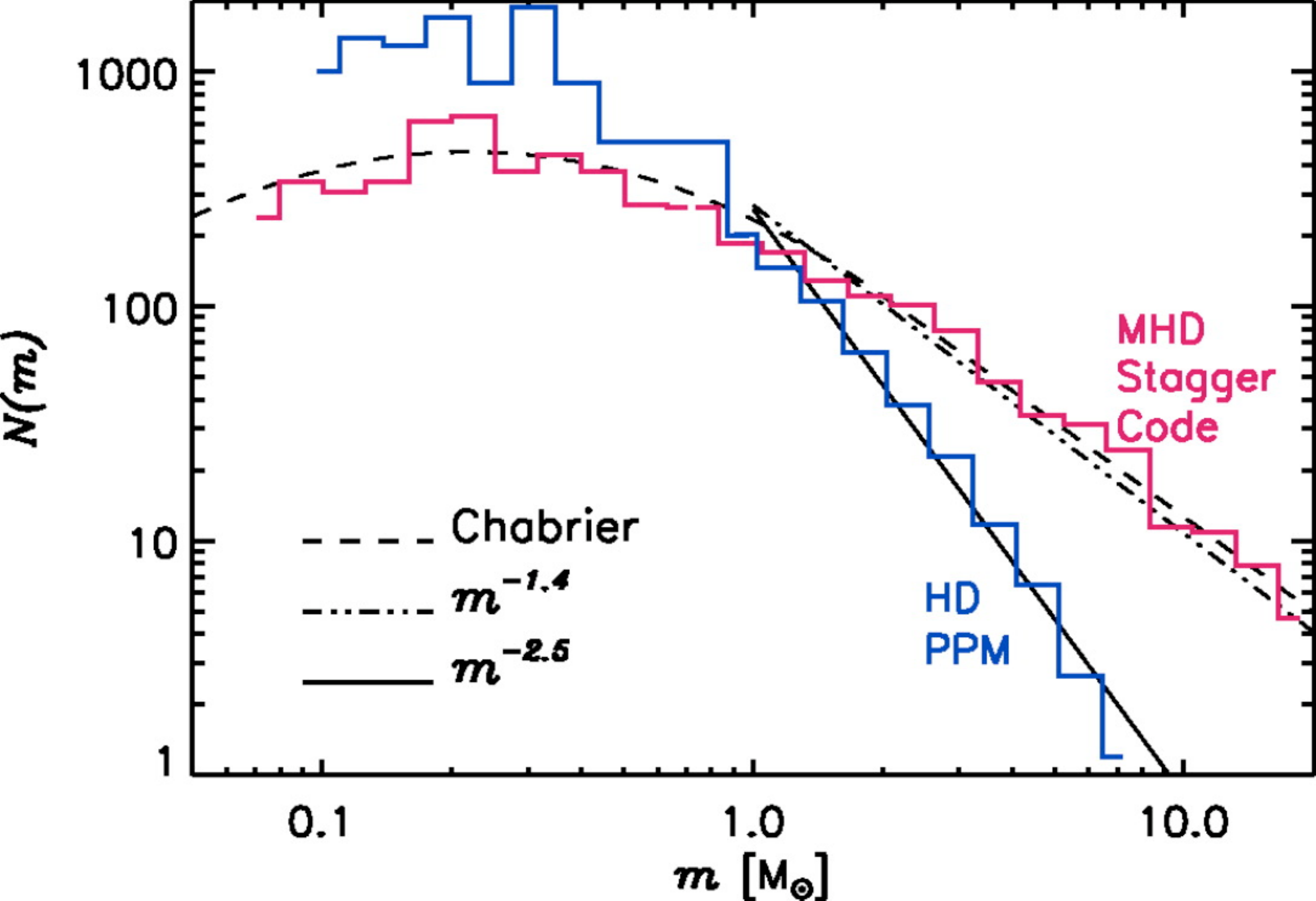}
  \caption{The distribution of core masses produced in a simulations of turbulence, using hydrodynamics (blue) and magnetohydrodynamics (red). The overplotted dashed line shows the IMF, using the functional form of \cite{chabrier03a}. Reprinted with permission from the AAS from \cite{padoan07a}.
  \label{fig:padoanimf}}
\end{figure}

To make this more quantitative, recall our result that the distribution of densities inside a molecular cloud, and thus this distribution of pressures (since $P\propto \rho$ in an isothermal gas), is lognormal. Thus the number of stars formed at a given mass is given by the number of fragments of that mass produced by the turbulence multiplied by the probability that each fragment generated is bound:
\begin{equation}
\xi(m) = \frac{dn_{\rm frag}}{dm} \int_{P_{\rm min(m)}}^\infty \frac{dp}{dP} \,dP,
\end{equation}
where $P_{\rm min}$ is the minimum pressure required to make an object of mass $m$ unstable, and $dP/dp$ is the (lognormal) distribution of pressures. The effect of this integral is to impose a lognormal turndown on top of the powerlaw produced by turbulence. Simulations seem to show fragments forming in a manner and with a mass distribution that is in good agreement with this model (Figure \ref{fig:padoanimf}).

\paragraph{The Evolution of Massive Cores}

By itself this model is not complete, because it doesn't explain why the massive cores don't fragment further as they collapse. After all, a 1 $\msun$ core may only be about 1 Bonnor-Ebert mass, but a 100 $\msun$ core is 100 Bonnor-Ebert masses, so why doesn't it fragment to produce 100 small stars instead of 1 big one? Even for low mass cores there tends to be too much fragmentation in simulations, resulting in an overproduction of brown dwarfs compared to what we see.

\begin{figure}
\includegraphics[height=.3\textheight]{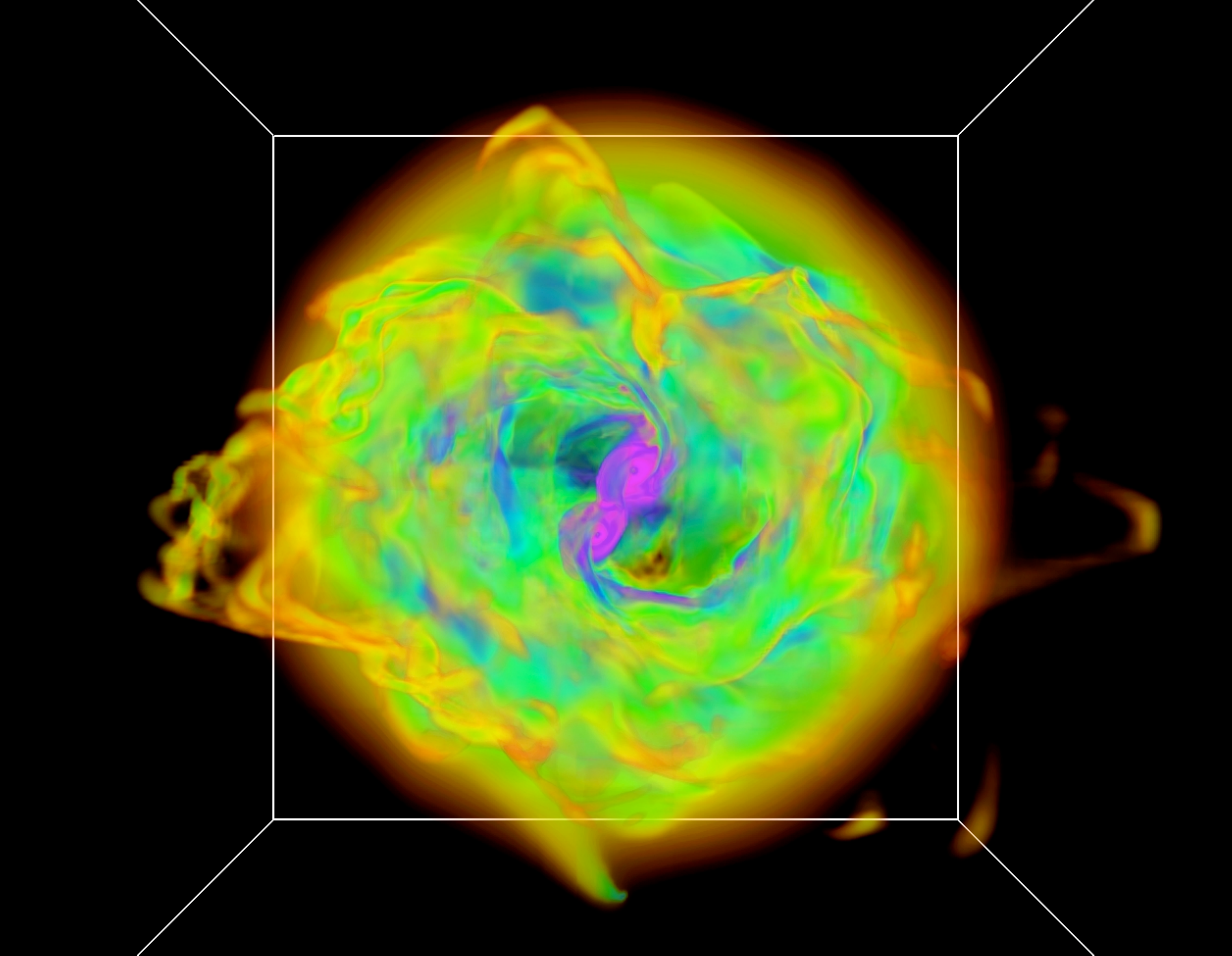}\includegraphics[height=.3\textheight]{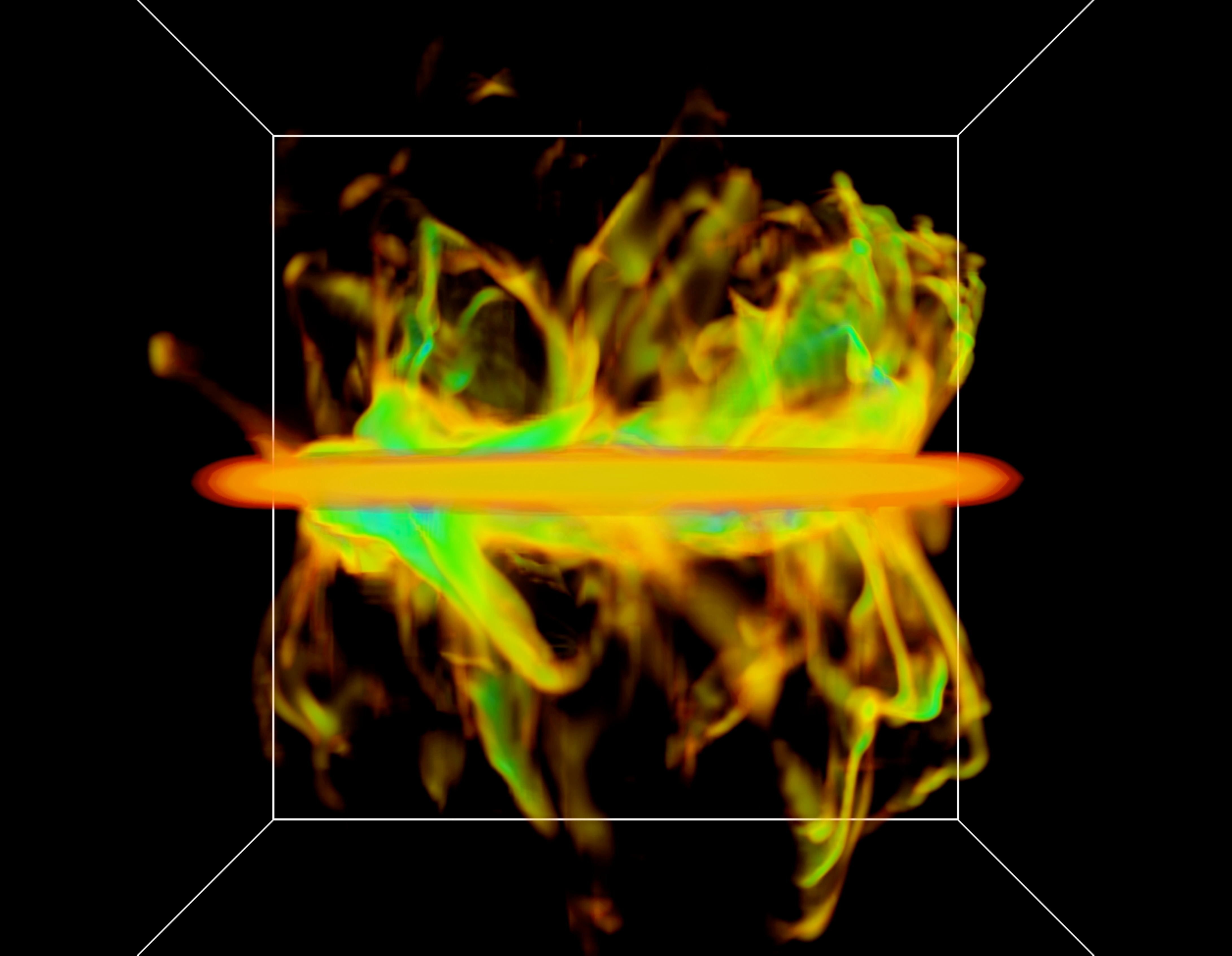}
  \caption{A volume rendering of the gas density a simulation of a the formation of a massive binary star system, showing the accretion disk face-on (left) and edge-on (right). Notice the Rayleigh-Taylor fingers that channel gas onto the accretion disk. Reprinted with permission from \cite{krumholz09c}.
  \label{fig:radrt}}
\end{figure}

The answer to that problem was provided in a series of papers by Krumholz et al.\ \cite{krumholz06b, krumholz07a, krumholz08a, krumholz10a}. Inside a collapsing core a first low mass star will form, and as matter accretes onto it, the accreting matter releases its gravitational potential energy as radiation. This is a {\it lot} of radiation. In a massive core the velocity dispersions tend to be supersonic, and the corresponding accretion rates $\sim \sigma^3/G$ are large, perhaps $\sim 10^{-4}-10^{-3}$ $\msun$ yr$^{-1}$. If one drops mass at this rate onto a protostar of mass $M$ and radius $R$, the resulting luminosity is
\begin{equation}
L = \frac{G M\dot{M}}{R} = 3\times 10^3\lsun \left(\frac{M}{\msun}\right) \left(\frac{\dot{M}}{10^{-4}\,\msun\mbox{ yr}^{-1}}\right) \left(\frac{R}{\rsun}\right)^{-1}.
\end{equation}

With a source of this luminosity shining from within it, a massive core is no longer isothermal. Instead, its temperature rises, raising the sound speed and suppressing the formation of small stars -- recall that, in an isothermal gas, $M_{\rm BE} \propto c_s^3 \propto T^{3/2}$. Then the problem becomes a dynamical one, in which there is a competition between secondary fragments trying to collapse and the radiation from the first object trying to raise their temperature and pressure to disperse them. One can study this result using radiation-hydrodynamic simulations, and the result is that, for sufficiently dense massive cores, the heating tends to win, and massive cores tend to form binaries, but fragment no further. (This latter point is good, because essentially all massive stars are observed to be binaries.)

A final difficulty with massive cores is radiation pressure. A cartoon version of the problem can be understood as follows. Massive stars have very short Kelvin times, so they will reach the main sequence and begin hydrogen burning while they are still accreting. Now consider the force per unit mass exerted by the star's radiation on the gas around it. This is
\begin{equation}
f_{\rm rad} = \frac{\kappa L}{4\pi r^2 c},
\end{equation}
where $\kappa$ is the opacity per unit mass. We can compare this to the gravitational force per unit mass
\begin{equation}
f_{\rm grav} = \frac{G M}{r^2}
\end{equation}
to form the Eddington ratio
\begin{equation}
f_{\rm Edd} = \frac{f_{\rm rad}}{f_{\rm grav}} = \frac{\kappa}{4\pi G c} \left(\frac{L}{M}\right) 
= 3.8\times 10^{-4} \left(\frac{\kappa}{5\mbox{ cm}^2\mbox{ g}^{-1}}\right) \left(\frac{L}{M}\right)_{\odot},
\end{equation}
where the value of $\kappa$ we've plugged in is typical for dusty interstellar gas absorbing near-IR photons. Thus we expect radiation force to exceed gravitational force once the star has a light to mass ratio larger than a few thousand in Solar units. This happens at a mass $M\sim 20$ $\msun$.

So how can bigger stars form, when they should repel rather than attract interstellar matter? This is a classic problem in star formation, and it led to all sorts of exotic theories for how massive stars form, e.g.\ that they form via stellar collisions in dense clusters. If any of these models are right, then the picture we've just outlined cannot be correct. Fortunately, it turns out that there is a more prosaic answer. Real life is not spherically symmetric, and using radiation to try to hold up infalling gas proves to be an unstable situation. The instability is not all that different from garden variety Rayleigh-Taylor instability, with radiation playing the role of the light fluid (Figure \ref{fig:radrt}).

%%%%%%%%%%%%%%%%%%%%%%%%%%%%%%%%%%%%%%%%%%%%%%%%
%% BACKMATTER
%%%%%%%%%%%%%%%%%%%%%%%%%%%%%%%%%%%%%%%%%%%%%%%%

\begin{theacknowledgments}
MRK is supported by an Alfred P. Sloan Fellowship; the US National science Foundation through grants AST-0807739 and CAREER-0955300; and NASA through Astrophysics Theory and Fundamental Physics grant NNX09AK31G and a Spitzer Space Telescope Cycle 5 Theoretical Research Program grant.
\end{theacknowledgments}



\bibliographystyle{aipproc}   % if natbib is available

\bibliography{refs}

\IfFileExists{\jobname.bbl}{}
 {\typeout{}
  \typeout{******************************************}
  \typeout{** Please run "bibtex \jobname" to optain}
  \typeout{** the bibliography and then re-run LaTeX}
  \typeout{** twice to fix the references!}
  \typeout{******************************************}
  \typeout{}
 }

\end{document}